\documentclass[aps,prd,showkeys,twocolumn,amsfonts,amssymb,amsmath,nofootinbib,letterpaper,eqsecnum,longbibliography]{revtex4-2}
\pdfoutput=1
\usepackage{graphicx}
\usepackage[T1]{fontenc}
\usepackage[dvipsnames]{xcolor}
\usepackage{blindtext}
\usepackage{physics}
\usepackage{hyperref}
\usepackage{mathtools}
\usepackage{bm}
\usepackage{bigints} %
\usepackage[normalem]{ulem} %
\usepackage{manfnt} %
\usepackage[english]{babel}

\usepackage{booktabs}

\usepackage{scalerel}
\usepackage{tikz}
\usetikzlibrary{svg.path}

\definecolor{orcidlogocol}{HTML}{A6CE39}
\tikzset{
  orcidlogo/.pic={
    \fill[orcidlogocol] svg{M256,128c0,70.7-57.3,128-128,128C57.3,256,0,198.7,0,128C0,57.3,57.3,0,128,0C198.7,0,256,57.3,256,128z};
    \fill[white] svg{M86.3,186.2H70.9V79.1h15.4v48.4V186.2z}
                 svg{M108.9,79.1h41.6c39.6,0,57,28.3,57,53.6c0,27.5-21.5,53.6-56.8,53.6h-41.8V79.1z M124.3,172.4h24.5c34.9,0,42.9-26.5,42.9-39.7c0-21.5-13.7-39.7-43.7-39.7h-23.7V172.4z}
                 svg{M88.7,56.8c0,5.5-4.5,10.1-10.1,10.1c-5.6,0-10.1-4.6-10.1-10.1c0-5.6,4.5-10.1,10.1-10.1C84.2,46.7,88.7,51.3,88.7,56.8z};
  }
}

\newcommand\orcidlink[1]{\href{https://orcid.org/#1}{\mbox{\scalerel*{
\begin{tikzpicture}[yscale=-1,transform shape]
\pic{orcidlogo};
\end{tikzpicture}
}{|}}}}

\usepackage{hyperref}

\allowdisplaybreaks[4]

\makeatletter

\newcommand{\ringplus}{\mathbin{\text{\@ringplus}}}

\newcommand{\@ringplus}{%
  \ooalign{\hidewidth\raise1.3ex\hbox{\tiny$\circ$}\hidewidth\cr$\m@th+$\cr}%
}

\newcommand{\ringminus}{\mathbin{\text{\@ringminus}}}

\newcommand{\@ringminus}{%
  \ooalign{\hidewidth\raise0.9ex\hbox{\tiny$\circ$}\hidewidth\cr$\m@th-$\cr}%
}
\makeatother

\newcommand{\tp}[0]{\mathrm{T}}

\newcommand{\cube}{\text{\mancube}}

\let\originalleft\left
\let\originalright\right
\renewcommand{\left}{\mathopen{}\mathclose\bgroup\originalleft}
\renewcommand{\right}{\aftergroup\egroup\originalright}

\DeclareFontFamily{U}{wncy}{}
\DeclareFontShape{U}{wncy}{m}{n}{<->wncyr10}{}
\DeclareSymbolFont{mcy}{U}{wncy}{m}{n}
\DeclareMathSymbol{\Sh}{\mathord}{mcy}{"58}

\newcommand{\negspace}{\!}

\newcommand{\lsub}[2]{{\protect\vphantom{#1}}_{#2} \negspace {#1}}
\newcommand{\rsub}[2]{{#1} \negspace {\protect\vphantom{#1}}_{#2}}

\newcommand{\ketsub}[2]{\rsub {\ket{#1}} {#2}}
\newcommand{\brasub}[2]{\lsub {\bra{#1}} {#2}}

\newcommand{\inprod}[2]{\left\langle {#1} | {#2} \right\rangle}

\providecommand{\abs}[1]{\left\lvert{#1}\right\rvert}
\newcommand{\abss}[1]{\lvert{#1}\rvert}

\newcommand{\reals}[0]{\mathbb{R}}

\renewcommand{\vec}[1]{\bm{\mathrm{#1}}}

\usepackage{graphicx}
\graphicspath{{./figures/}}

\renewcommand{\vec}[1]{\bm{\mathbf{#1}}}

\newcommand{\sonic}{{\mathrm{s}}}
\newcommand{\particle}{{\mathrm{p}}}

\newcommand{\init}{{\mathrm{i}}}
\newcommand{\final}{{\mathrm{f}}}

\newcommand{\absvec}[1]{\left\vert{\vec{#1}}\right\vert}

\newcommand{\vacsonicket}{\ketsub 0 \sonic}
\newcommand{\vacsonicbra}{\brasub 0 \sonic}

\newcommand{\charge}{{g}}

\hypersetup{colorlinks=true,urlcolor=MidnightBlue,linkcolor=black,citecolor=MidnightBlue,filecolor=black}

\begin{document}

	\title{Particle scattering in a sonic analogue of special relativity}

	\author{Scott L. Todd\,\orcidlink{0000-0002-0687-9437}}
	\email{scott@todd.science}
	\author{Giacomo Pantaleoni\,\orcidlink{0000-0002-8688-0656}}
	\author{Valentina Baccetti\,\orcidlink{0000-0002-5680-1574}}
	\author{Nicolas C. Menicucci\,\orcidlink{0000-0002-3964-233X}}
\affiliation{%
	Center for Quantum Computation and Communication, RMIT University,\\ %
	Melbourne, Victoria 3000, Australia\\
}
\date{February 11, 2022}

	\begin{abstract}
		We investigate a simple toy model of particle scattering in the flat spacetime limit of an analogue-gravity model. The analogue-gravity medium is treated as a scalar field of phonons that obeys the Klein-Gordon equation and thus admits a Lorentz symmetry with respect to $c_\sonic$,
		the speed of sound in the medium. The particle from which the phonons are scattered is external to the system and does not obey the sonic Lorentz symmetry that the phonon field obeys. In-universe observers who use the exchange of sound to operationally measure distance and duration find that the external particle appears to be a sonically Lorentz-violating particle. By performing a sonic analogue to Compton scattering, in-universe observers can determine if they are in motion with respect to their medium. If in-universe observers were then to correctly postulate the dispersion relation of the external particle, their velocity with respect to the medium could be found.
	\end{abstract}

\maketitle

\section{Introduction}

Analogue-gravity models provide an indirect way to probe the physics of gravitational systems for which the actual (i.e.,~nonanalogue) experiments are currently inaccessible. Perhaps most well known are the acoustic analogues to black holes (also called \textit{dumb holes}\footnote{``Dumb,'' in this case, being a synonym for ``mute.''}), originally proposed by Unruh~\cite{unruhExperimentalBlackHoleEvaporation1981} to provide a theoretical and experimental testbed for the study of Hawking radiation~\cite{hawkingBlackHoleExplosions1974,hawkingParticleCreationBlack1975} in a system where the microscopic physics is understood. Subsequently, analogue models for a variety of general relativistic and semiclassical gravitational phenomena have been discovered, and many such models are now being experimentally realized---notably, acoustic analogues for Hawking radiation~\cite{weinfurtnerMeasurementStimulatedHawking2011,steinhauerObservationSelfamplifyingHawking2014,steinhauerObservationQuantumHawking2016,munozdenovaObservationThermalHawking2019}, optical-media analogues for Hawking radiation~\cite{droriObservationStimulatedHawking2019}, and analogues for cosmological expansion and particle production~\cite{eckelRapidlyExpandingBoseEinstein2018} have all successfully seen experimental realization to date. A comprehensive list of analogue-gravity models and the research into them as of 2011 can be found in the extensive review article by Barcel\'{o}, Liberati, and Visser and the references therein~\cite{barceloAnalogueGravity2011}, while references to some noteworthy experimental results in the interim can be found in a short article by Jacquet, Weinfurtner, and K\"{o}nig~\cite{jacquetNextGenerationAnalogue2020a}, which discusses both the history and future of experiments within the analogue gravity research endeavor. The aforementioned article by Jacquet {\it et al.} brings attention to experimental work on superradiant scattering by Torres {\it et al.}~\cite{torresRotationalSuperradiantScattering2017,torresQuasinormalModeOscillations2020}, as well as to developments on modeling gravitational phenomena within superconducting circuits~\cite{blencoweAnalogueGravitySuperconducting2020} (see Tian {\it et al.}~\cite{tianAnalogCosmologicalParticle2017,tianAnalogueHawkingRadiation2019a} and Lang and Sch\"{u}tzhold~\cite{langAnalogCosmologicalParticle2019} for additional work on superconducting circuits within the context of analogue gravity). While by no means extensive, a selection of noteworthy theoretical results since 2011 include~\cite{goulartHiddenGeometriesNonlinear2011,finazziCosmologicalConstantLesson2012,zubkovEmergentGravityGraphene2015,liberatiAnalogueGravityModels2017,leonhardtClassicalAnalogUnruh2018,fiferAnalogCosmologyTwofluid2019}.

Studying relativistic phenomena with analogue models is, of course, only a valid approach provided that the analogue system can be faithfully mapped back to the actual physical system of interest. One obvious---and seemingly detrimental---way in which the mapping between analogue models and real physical systems of interest seems to fail is that analogue models are not truly relativistic. In the most obvious case, there exists a preferred reference frame: the rest frame of the analogue-gravity medium itself. Previous work by Barcel\'{o} and Jannes~\cite{barceloRealLorentzFitzGeraldContraction2008} has highlighted that the natural way to consider analogue-gravity systems as genuine relativistic analogues is to use devices and observers that are \textit{internal} to the analogue medium  (e.g.,~they can be constructed from quasiparticle excitations of the medium itself) to make internal measurements of phenomena within the analogue universe. Phenomena like the Lorentz-FitzGerald contraction can be operationally shown to appear in a relativistically reciprocal manner by utilizing this notion of internal devices and observers: internal observers who are at rest with respect to the medium will \textit{measure} moving observers' devices to be length contracted (they are), and internal observers who are moving with respect to the medium will \textit{measure} stationary observers' devices to be length contracted (even though they are not)~\cite{toddSoundClocksSonic2017}. We call such observers \textit{in-universe observers}.

One way to understand the operational emergence of relativity from an analogue-gravity model is to consider the flat-spacetime limit: for an acoustic system, the flat-spacetime limit is exactly a sonic analogue to Lorentz ether theory,\footnote{Provided that one only considers the motion of sound waves belonging to the linear part of the medium's dispersion relation.} and it is known that Lorentz ether theory and special relativity are observationally equivalent~\cite{bellHowTeachSpecial1976,szaboLorentzianTheoriesVs2011}.
Therefore, from the operational viewpoint that considers only the measurements made by in-universe observers~\cite{toddSoundClocksSonic2017}, the sonic analogue to Lorentz ether theory can be seen to faithfully map to a sonic analogue of special relativity, which we call \textit{sonic relativity}. For the purposes of this paper, our considerations will be restricted to such sonically relativistic analogues of special relativity.

In attempting to utilize analogue-gravity models as a means to study phenomena within the overlap of relativistic physics and quantum physics, it would prove useful to understand how operational measurements of quantum mechanical phenomena can be performed in such systems. In the real world, quantum field theory (QFT) is the most well-developed theory incorporating the effects of special relativity and quantum mechanics, and within the context of QFT, scattering experiments are one of our most valuable experimental tools for probing the dynamical interactions of physical phenomena.\footnote{See almost any textbook on QFT for a discussion on scattering and its relevance to experiments, e.g.,~\cite{peskinIntroductionQuantumField2018a,srednickiQuantumFieldTheory2007,schwartzQuantumFieldTheory2014,lancasterQuantumFieldTheory2014}. For a more in-depth discussion on the nature of detectors themselves, see~\cite{leoTechniquesNuclearParticle2012,grupenHandbookParticleDetection2012}.} To this end---and as the title of this paper suggests---our focus within this paper shall be on scattering experiments rather than on other types of experiments that can be conducted within the framework of QFT.

If we are to attempt to utilize analogue-gravity models to their full potential (i.e.,~as a method by which to carry out actual experiments), then we should---as a first step---seek to describe and understand those detector models that can conceivably be experimentally realized and that can be used in conjunction with analogue-gravity systems. This leads to several questions. For example, what constitutes an appropriate model of a particle detector within an analogue-gravity model? How much of our understanding of quantum detectors from actually relativistic theories can we apply to analogue-gravity systems? If we consider deviations away from the idealized case in which we only consider that which is \textit{internal} to an analogue-gravity model, how is our understanding of detector models impacted?\footnote{Where the notion of \textit{internal} is as per the discussions of Barcel\'{o} and Jannes~\cite{barceloRealLorentzFitzGeraldContraction2008}.} In particular, this last question provides the basis of the work presented in this paper.

We will herein consider scattering experiments within analogue-gravity models. We do so in keeping with the philosophy of a previous paper that was published by two of the present authors~\cite{toddSoundClocksSonic2017}: we consider the measurements made by in-universe observers within an analogue-gravity system, and we restrict these observers to
measuring duration and distance solely through the exchange of sound pulses. This ensures their measuring devices obey sonic Lorentz symmetry.
We assume that in-universe observers have access to the apparatus required to do scattering experiments with phonons. That is, they can produce phonons, and they can detect recoiling phonons through a ``click'' of a particular detector within an array of such detectors. We neglect the detailed questions of how such items are constructed, understanding that our assumptions are enough to meaningfully discuss scattering experiments in this setting.\footnote{For the reader who is nevertheless curious about the possibility of actually realizing in-universe detectors, a sensible-seeming starting point would be to consider devices whose operation utilizes optomechanical interactions. To this end, recent work involving Brillouin scattering~\cite{eggletonInducingHarnessingStimulated2013,kittlausOnchipIntermodalBrillouin2017,eggletonBrillouinIntegratedPhotonics2019, Gertler_2020} and recent work involving surface acoustic waves (SAWs) coupled to qubits (to yield qubit-SAW devices)~\cite{satzingerQuantumControlSurface2018,bienfaitPhononmediatedQuantumState2019} may provide useful food for thought. Graphene and topological semimetals are yet another possibility~\cite{zubkovEmergentGravityGraphene2015,Guan:2017up}.}
We seek to determine the results that in-universe observers measure of phonon scattering experiments, specifically in the case that phonons are scattered from an external particle---that is, one whose dispersion relation is nonrelativistic.

The contents of this paper are as follows: in Sec.~\ref{Sec:In-universe observers} we review the notion of in-universe observers~\cite{toddSoundClocksSonic2017}. In Sec.~\ref{Sec:Paper_Strategy} we elucidate further on our aims, provide a sketch of the approach that is used throughout this paper, state the conventions that we shall use, and give a schematic of the scattering events experiments that we consider. In Sec.~\ref{Sec:PhononScattering} we determine the kinematic expressions of a toy model of phonon scattering from two types of particles: one type of particle is {sonically Lorentz obeying} (internal particle), and the other is {sonically Lorentz violating} (external particle). In both cases, we first obtain the kinematic description of scattering in the {laboratory frame} and then in the frame of an in-universe observer who is comoving with the particle from which the phonons scatter. (This is the approach laid in out in Sec.~\ref{Sec:Paper_Strategy}.) In Sec.~\ref{Sec:Scattering cross sections} we consider phonon scattering from a first-quantized particle with a Newtonian dispersion relation (an external particle) and present the results that demonstrate that these scattering experiments reveal information about the existence of a preferred rest frame to in-universe observers. In Sec.~\ref{Sec:Discussion} we explicitly explain in what way the results presented in the previous section reveal information about the existence of a preferred rest frame. In Sec.~\ref{Sec:Conclusion} we summarize our results and demonstrate that the Standard-Model extension might provide a way to model phonon scattering from external particles as described by in-universe observers.

\section{In-universe observers}
\label{Sec:In-universe observers}

	Assume that there are in-universe observers in an analogue-gravity universe who believe in the principle of special relativity, with the exception that sound takes the place of light~\cite{toddSoundClocksSonic2017}. So defined, in-universe observers can be constructed using classical (nonquantum) rulers and clocks, as in~\cite{toddSoundClocksSonic2017}\footnote{In-universe observers could also be constructed from quasiparticle excitations of the medium itself, as proposed in~\cite{barceloRealLorentzFitzGeraldContraction2008}. However, by their nature, quasiparticle excitations are not, in general, well localised in energy, momentum, or space. This implies that considering a classical rest frame for in-universe observers may not be correct. If one were to consider in-universe observers made from quasiparticle excitations, the correct procedure would be to use the framework of quantum reference fames, as formulated in~\cite{Giacomini:2019uf}. In order to avoid this additional layer of complexity we choose to work with classical in-universe observers as defined in~\cite{toddSoundClocksSonic2017}.}. Taking the coordinates $x$, $y$, $z$, and $t$ to be measured operationally by some in-universe observer via the exchange of sound, and defining $\vec{x}=(x,y,z)$ as a notational shorthand, we can construct the following four-vector
	\begin{equation}
	X^\mu\coloneqq \begin{pmatrix}
	c_\sonic t\\
	\vec{x}
	\end{pmatrix}.
	\label{Eq:Four position}
	\end{equation}
	This four-vector is a sonically Lorentz covariant object, and in keeping with special relativity we denote this the sonic four-position. Of course, there is nothing special about Cartesian coordinates: the sonically Lorentz covariant nature of the sonic four-position is true for any set of orthogonal coordinates. We have merely made the choice to pick Cartesian coordinates for the purposes of demonstration.

	Consider now two sets of in-universe observers who are initially traveling in the same direction, which we shall call the $z$ direction, by convention, with respect to their sound-carrying medium. The first set of observers are traveling with velocity $\vec v_1 =  (0, 0, v_1)$ with respect to the medium, and the second set of observers are traveling with a velocity of $\vec v_2 = (0, 0, v_2)$ with respect to the medium. Note that the operational measurements of distance and duration allow for in-universe observers to operationally calculate velocities~\cite{toddSoundClocksSonic2017}. In the particular case that we are considering, the operationally determined sonic fractional velocity of the second set of observers as measured by the first set of observers will be given (in terms of the laboratory-frame values of quantities) by the following expression:
	\begin{equation}
	\beta=\dfrac{\beta_2-\beta_1}{1-\beta_2 \beta_1},
	\label{Eq:Relativistic velocity composition}
	\end{equation}
	where $\beta_1=v_1/c_\sonic$ and $\beta_2=v_2/c_\sonic$. The quantity $\beta$ is the \textit{boost parameter} of the second frame with respect to the first frame.

	Each set of observers can describe the sonic four-position from their own operational measurements of distance and duration, as discussed above. Denote $X^\mu$ to be the components of the sonic four-position as operationally determined by the first set of observers, and ${X^\prime}^\nu$ to be the components of the sonic four-position as operationally determined by the second set of observers. The first set of observers can relate the components ${X^\prime}^\nu$ to $X^\mu$ with the following familiar equation:
	\begin{equation}
	\label{eq:LorentzBoost}
	{X^\prime}^\nu = \Lambda^\nu_{~\mu}X^\mu.
	\end{equation}
	For the particular case that we have been discussing $\Lambda^\nu_{~\mu}$ is given by\footnote{Note that this definition is \textit{not} simply some vacuous one that we merely assert to be true axiomatically. While we do not offer a derivation within this paper, it is indeed possible to take the content of \cite{toddSoundClocksSonic2017} and formulate a proper four-vector description of coordinates from the operational measurements made by in-universe observers. In doing so, the Lorentz transformation matrix naturally arises as the transformation relating the four-position between different in-universe observer frames.}
	\begin{equation}
	\Lambda^\nu_{~\mu} =
	\begin{pmatrix}
	\gamma & 0 & 0 & -\beta\gamma\\
	0  & 1 & 0 & 0\\
	0 & 0 & 1 & 0\\
	-\beta\gamma & 0 & 0 & \gamma\\
	\end{pmatrix}
	,
	\label{Eq:LorentzBoostExplicit}
	\end{equation}
	where $\beta$ is the operationally determined sonic fractional velocity of the second set of observers with respect to the first set of observers, given by Eq.~\eqref{Eq:Relativistic velocity composition}, and ${\gamma\coloneqq 1/\sqrt{1-\beta^2}}$.

	Another way to see that measurements of duration and distance made using only the exchange of sound signals can be grouped together into a Lorentz covariant object, $X^\mu$, is that the object being used to define spatial and temporal measurements (the phonon) obeys, in our simplified toy model, the Klein-Gordon equation~\cite{toddSoundClocksSonic2017}. The Klein-Gordon equation admits a Lorentz symmetry, and as a consequence of this, any measurements that are made using phonons inherit the Lorentz symmetry that the phonons themselves obey.

	A consequence of the existence of the sonic four-position vector and its associated Lorentz transformation is that all physical quantities whose values can be determined geometrically---that is to say, from the components of the sonic four-position---transform exactly as expected from special relativity. For our purposes, we note that measurements of spatial and/or temporal coordinates (i.e.,~the components $X^\mu$ or ${X^\prime}^\nu$) can be used by in-universe observers to operationally determine the wavelength and frequency of sound waves, and also as a way to operationally determine angles: as a result, between any pair of in-universe observer reference frames the wavelengths and frequencies of sound waves obey the relativistic Doppler shift formula, and the angle of propagation of an acoustic ray will change via the relativistic aberration formula (where, in both cases, the speed of sound replaces the speed of light). The relativistic Doppler shift formula and the relativistic aberration formula are given, respectively, by:
	\begin{align}
	\omega &= \gamma\left(1+\beta\cos\theta^\prime\right)\omega^\prime \label{Eq:Doppler_shift}\\
	\cos\theta &= \dfrac{\cos\theta^\prime+\beta}{1+\beta\cos\theta^\prime}, \label{Eq:Aberration}
	\end{align}
    where the last relation leads to the following useful expression,
    \begin{equation}
    (1 - \cos \theta)=
    \frac{1-\beta}{1 + \beta \cos \theta'}
    (1- \cos \theta')
     \label{Eq:oneminuscostheta}
    ,
	\end{equation}
	and where in all cases $\beta$ is the boost parameter of the second frame with respect to the first [given by Eq.~\eqref{Eq:Relativistic velocity composition}], an \textit{unprimed} symbol denotes the in-universe observer measured value of some physical quantity in the first frame, and a \textit{primed} symbol denotes the value of that same physical quantity as measured by in-universe observers in the second frame. Note that in the specific case that we are considering, the angle $\theta$ is measured relative to the $z$ axis and the angle $\theta^\prime$ is measured relative to the $z^\prime$ axis.

	For notational brevity in what follows, define the \textit{Doppler factor} to be
	\begin{equation}
	    D\coloneqq\sqrt{\dfrac{1+\beta}{1-\beta}}.
	    \label{Eq:Doppler_factor}
	\end{equation}
	In the particular case that $\theta^\prime=0$, the Doppler factor can be used to simplify Eq.~\eqref{Eq:Doppler_shift} to	$\omega=D\omega^\prime$.

	\section{Aim and Approach}
	\label{Sec:Paper_Strategy}

	\subsection{Aim}

	Our purpose is to investigate a particular type of scattering experiment within the context of analogue-gravity models. In particular, we choose to investigate a scattering process that is in some sense analogous to Compton scattering. In true Compton scattering~\cite{comptonQuantumTheoryScattering1923,peskinIntroductionQuantumField2018a}, a photon is scattered from a charged particle. By analogy we choose to analyze the scattering of phonons within an analogue-gravity system. We consider phonon scattering from two different types of particle, which we label as either \textit{sonically Lorentz-obeying} or \textit{sonically Lorentz-violating} based on their dispersion relation.

	Our aim is the following: characterize the in-universe observer perspective of these scattering experiments, with the specific goal of determining what in-universe observers can learn from the results of scattering experiments involving sonically Lorentz-violating particles, including what they can learn about their state of motion with respect to the medium.

	\subsection{Sketch of our approach}

	In a typical derivation of scattering within quantum field theory, one often makes use of relativistic arguments to simplify the derivation by, for example, moving into the center-of-mass frame of the system~\cite{peskinIntroductionQuantumField2018a}. While such an approach indeed simplifies the derivation for phonon scattering from sonically Lorentz-obeying particles~\cite{barceloRealLorentzFitzGeraldContraction2008}, this is certainly not the case for phonon scattering from sonically Lorentz-violating particles. In this paper we take the following approach to our derivations:
	\begin{enumerate}
	    \item We anchor ourselves to the laboratory frame, which is the frame in which the analogue-gravity medium is assumed to be at rest. In this particular frame, we know the actual dynamics, and we know that energy and momentum must be conserved.
	    \item  We identify, in the laboratory frame, the dispersion relations that apply to the phonon and to the particle from which the phonon will scatter.
	    \item Utilizing the relevant dispersion relations, we obtain the laboratory-frame kinematic expression for scattering by insisting on energy and momentum conservation.
	    \item With the laboratory-frame kinematics determined, we utilize the appropriate transformation rules to obtain the kinematic description of scattering from the perspective of the in-universe observer frame that is comoving with respect to the particle prior to phonon scattering.
	\end{enumerate}
	This procedure grounds our calculation in the frame for which we know the dynamics: the laboratory frame. We do all of our dynamical calculations from this frame. The final step then determines how this behavior would appear to an in-universe observer who perceives the particle to be initially at rest in their own frame (even though both the observer and the particle may in fact be moving with respect to the medium). This allows us to tease out the differences between experiments that look the same to in-universe observers---particle initially at rest in the observer's frame---but that \textit{actually} differ in their initial velocity with respect to the analogue-gravity medium---i.e., with respect to the \textit{sonic ether}.

	As a sanity check, we first utilize this approach to demonstrate that the kinematic description of phonon scattering from a sonically Lorentz-obeying particle shows no dependence on the initial motion with respect to the medium, as would be the case for ordinary scattering in a fully Lorentz-obeying model~\cite{barceloRealLorentzFitzGeraldContraction2008} and in ordinary QFT~\cite{peskinIntroductionQuantumField2018a}.	We then proceed to use the same approach to arrive at a kinematic description for phonon scattering from a sonically Lorentz-violating particle.

	A mathematical sketch of this approach is the following. The energy and momentum conservation in the laboratory frame is simply:\footnote{Note that we have restricted our considerations to the linear part of the medium's dispersion relation through our particular choices for the phonon's energy ($\hbar\omega$) and momentum ($\hbar\mathbf{k}$).}
	\begin{align}
	    E_\init + \hbar\omega_\init &= E_\final + \hbar\omega_\final,\\
	    \vec{p}_\init + \hbar\vec{k}_\init &= \vec{p}_\final + \hbar\vec{k}_\final,
	    \label{Eq:ConservationEnergyMomentumGeneral}
	\end{align}
	where the notation, which is used throughout, is defined in Table~\ref{Tab:Physical_meaning_of_quantities}. The dispersion (energy-momentum) relations allow us to express energies as functions of momenta:
	\begin{align}
	    E&=E(\vec{p}),
	    &
	    \hbar\omega&=\hbar\omega(\vec{k}),
	\end{align}
	from which we can rewrite the conservation of energy and momentum like
	\begin{align}
	    E_\init(\vec{p}_\init) + \hbar\omega_\init(\vec{k}_\init) &= E_\final(\vec{p}_\final) + \hbar\omega_\final(\vec{k}_\final),\label{Energy_conservation_with_functions_of_momenta}\\
	    \vec{p}_\final &= \vec{p}_\init - \hbar\left(\vec{k}_\final - \vec{k}_\init\right).\label{Eq:Final_momentum_from_momentum_conservation}
	\end{align}
	From here, we explicitly substitute our dispersion relations into Eq.~\eqref{Energy_conservation_with_functions_of_momenta}, using the conservation of momentum [Eq.~\eqref{Eq:Final_momentum_from_momentum_conservation}] in the explicit functional form of $E_\final(\vec{p}_\final)$. (The specific functional forms that we will use will be given in subsequent sections.) Finally, we perform any valid algebraic steps that are required to obtain the desired kinematic expression for the Compton-like scattering of phonons. The resulting expression, by construction, obeys energy and momentum conservation, and is written entirely in terms of the the initial parameters of our system---which, in principle, we are free to control---and the final state of the phonon---which we can envisage experimentally detecting using some appropriate apparatus, e.g.,~a detector array comprised of some sonic analogue to photomultiplier tubes.

	Our final step is to utilize the laboratory-frame kinematic description of scattering---along with our understanding of how the values of physical quantities as measured by in-universe observers change between different reference frames---to determine the in-universe observer description of the kinematics. Specifically, we choose to obtain the kinematic description of scattering in the frame of an in-universe observer who is initially comoving with the particle from which the phonon will scatter---i.e.,~we are interested in the kinematic description as seen by the observer who initially believes the particle to be at rest. Before we can explicitly list the transformation rules for the values of physical quantities, it is important to recall a few key facts and to have a schematic of the scattering experiment in mind.

\begin{table}[t]
\centering
\begin{tabular}{c c c}
\toprule
~~Phonon~~
& ~~Particle~~
 & Meaning
 \\
\midrule
 \addlinespace[0.5ex]
\addlinespace[0.7ex]
$\hbar \vec k_\init$ & $\vec p_\init$ & Initial three-momentum (lab frame) \\
    \addlinespace[0.7ex]
$\hbar \vec k_\final$ & $\vec p_\final$ & Final three-momentum (lab frame) \\
    \addlinespace[0.7ex]
$k = \abs{\vec k}$ & $p = \abs{\vec p}$ & three-vector magnitude (lab frame)
\\
\bottomrule
\end{tabular}
    \caption{Notation for momenta of the phonon and particle. We restrict our initial setup to the case where $\vec k_\init$ and $\vec p_\init$ are in the positive $z$ direction.%
    }
    \label{Tab:Physical_meaning_of_quantities}
\end{table}

	\subsection{Schematic of our scattering experiment}

	Recall that the specific form of the transformation equations that we established in Sec.~\ref{Sec:In-universe observers} were predicated on the assumption that pairs of in-universe observer reference frames had coordinate axes that were aligned, and that the relative motion between pairs of in-universe observer reference frames was constrained to be in the $z$ direction. For the purposes of this paper, we choose only to consider experiments for which these assumption hold true. Also note that---as explicitly demonstrated in~\cite{toddSoundClocksSonic2017}---operational measurements of distance and duration that are made by in-universe observers who are stationary with respect to the analogue-gravity medium coincide with the equivalent measurements as made in the laboratory frame (up to unit conversions). That is to say, the laboratory frame \textit{is} an in-universe observer reference frame; it simply happens to be the in-universe observer reference frame corresponding to observers who are stationary with respect to the medium.
	\begin{figure}[tb]
		\includegraphics[width=\linewidth]{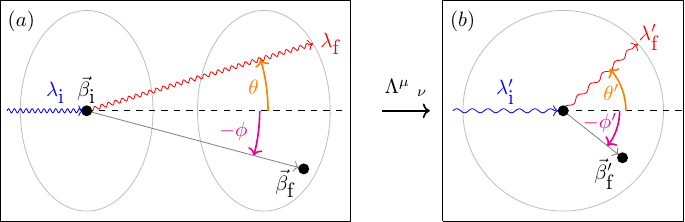}
		\caption{A scattering event as seen in (a) the laboratory frame and (b) the comoving in-universe observer frame. All quantities labeled in the figure are measured operationally with sound-based clocks and rulers~\cite{toddSoundClocksSonic2017} and thus their values as measured by in-universe observers transform with respect to a sonic Lorentz transformation. We envisage the particle from which scattering occurs to be centered with respect to a hollow shell of detectors that can detect phonons incident upon them. To in-universe observers in the comoving frame (b), the detector array appears to form a spherical shell defined by constant spatial coordinates; in the laboratory frame (a), the array of detectors appears to be Lorentz contracted in the direction of motion and the spatial coordinates defining this shell change between the scattering event and the detection event. The relativistic Doppler formula and the relativistic aberration formula relate the operationally measured values of wavelengths and angles between reference frames, respectively. The operationally measured value of the final velocity of the particle in different reference frames is related by the general form of the relativistic velocity composition formula~\cite{einsteinZurElektrodynamikBewegter1905} (see~\cite{einsteinElectrodynamicsMovingBodies} for an English translation).
		}
		\label{Fig:Compton Scattering not-boosted and boosted}
	\end{figure}

	With all of this in mind, we present in Fig.~\ref{Fig:Compton Scattering not-boosted and boosted} a schematic of the scattering experiments that we will consider throughout the remainder of this paper. We note that our scattering experiments are initialized such that the trajectories of our particles are parallel with respect to the incoming phonons; we \textit{define} these trajectories to point in the positive $z$ direction. Denoting velocities $\vec{v}=(v_x,v_y,v_z)$, we can write the initial velocity of our particle $\vec{v_\init}=(0,0,v_\init)$, where $v_\init\geq{0}$ by fiat of our assumptions; consequently, we have $\absvec{v_\init}=v_\init$. Denoting sonic fractional velocities to be $\vec{\beta}\coloneqq\vec{v}/c_\sonic$, we can write $\beta_\init\coloneqq\absvec{\beta_\init}=v_\init/c_\sonic$. Restricting ourselves to cases in which the particle's initial and final speeds are slower than $c_\sonic$, we always have that $\left\vert{\vec{\beta}}\right\vert<1$; taken in combination with $v_\init\geq{0}$, we have then that $0\leq{\beta_\init}<1$. The final velocity of the particle $\vec{\beta_\final}$ is allowed to have components in any direction, provided that $\left\vert{\vec{\beta_\final}}\right\vert<1$. As it turns out, though, we will ultimately be able to remove any reference to the final state of the particle (and thus to $\vec{\beta_\final}$) from the kinematic expressions that we obtain.

	We previously provided equations for the relativistic Doppler shift of sound waves [Eq.~\eqref{Eq:Doppler_shift}], and for the relativistic aberration formula [Eq.~\eqref{Eq:Aberration}]. As we have emphasized, in our particular scattering experiments we wish to consider the operational measurements made in the comoving in-universe observer frame. In the laboratory frame, we therefore have that $\beta=\beta_\init$---that is, the boost parameter of the comoving in-universe observer frame is numerically equal to the initial velocity of the particle, and thus the values of quantities in the laboratory frame (unprimed symbols) are related to the values of quantities in the comoving in-universe observer frame (primed symbols) in the following specific ways:
	\begin{align}
	\omega_\init
&
=
	\gamma_\init
	\left(
		1
	+
		\beta_\init
	\right)
	\omega^\prime_\init
=
    D_\init \omega^\prime_\init
    ,
\label{Eq:Transformation for initial frequency}
	\\
	\omega_\final
&
=
	\gamma_\init
		\left(
			1
		+
			\beta_\init
			\cos\theta^\prime
		\right)
		\omega^\prime_\final
		,
\label{Eq:Transformation for final frequency}
	\\
	\cos\theta
&
=
	\dfrac{
		\cos\theta^\prime
	+
		\beta_\init}{
		1
	+
		\beta_\init
		\cos\theta^\prime}
		,
\label{Eq:TransformationCosTheta}
	\\
	(1 - \cos\theta )
&
=
    \frac{1-\beta_\init}{1 + \beta_\init \cos \theta'}
    (1- \cos \theta')
    ,
    \label{Eq:TransformationOneMinusCosTheta}
	\end{align}
where Eqs.~\eqref{Eq:Transformation for final frequency}--%
\eqref{Eq:TransformationOneMinusCosTheta} follow directly from Eqs.~\eqref{Eq:Doppler_shift}--%
\eqref{Eq:oneminuscostheta}, respectively, and Eq.~\eqref{Eq:Transformation for initial frequency} follows from Eq.~\eqref{Eq:Doppler_shift} with ${\theta = \theta'=0}$ because the phonon is always initially traveling in the positive $z$ direction.

\section{Phonon scattering kinematics}
\label{Sec:PhononScattering}

    We now proceed to derive the kinematic expressions of Compton-like phonon scattering within an analogue-gravity model. As we stated in Sec.~\ref{Sec:Paper_Strategy}, we first obtain the kinematic expression for phonon scattering for the case of \textit{sonically Lorentz-obeying} particles (internal particles). Once we have obtained both the laboratory frame and comoving in-universe observer frame kinematics for phonon scattering from internal particles, we will then proceed to do the same for \textit{sonically Lorentz-violating} particles (external particles).

    \subsection{Phonon scattering from internal particles}
    \label{Sec:Phonon scattering from internal particles}

    \subsubsection{The laboratory frame kinematics of phonon scattering from internal particles}

    To provide some context, we imagine that, following Ref.~\cite{barceloRealLorentzFitzGeraldContraction2008}, internal particles are collective-excitation quasiparticles whose dynamical description is covariant with respect to the same sonic Lorentz symmetry that in-universe observer reference frames transform under. In other words, these quasiparticles are relativistic with respect to the speed of sound $c_\sonic$ of the analogue-gravity medium. Most obviously, one might imagine these particles to arise from the analogue-gravity medium itself (hence the use of the word ``internal''). However, there is no particular reason that these particles could not also arise in a separate medium with the same speed of sound, and to which the analogue-gravity medium is somehow coupled.

    For the purposes of this paper, we choose the following simple and familiar energy-momentum relations for our internal particles:
	\begin{align}
		E &= {\gamma}mc_\sonic ^2,
		&
		\mathbf{p} &= \gamma{m}\vec{v},
		\label{Eq:Relativistic_particle}
	\end{align}
    where here $\mathbf{v}$ is the velocity of the particle as measured in the laboratory frame, ${\gamma=1/\sqrt{1-\beta^2}}$, $\vec{\beta}=\vec{v}/c_\sonic$, and $\beta^2\coloneqq\vec{\beta}\cdot\vec{\beta}$. The parameter $m$ is the mass of the particle, and since the expressions for $E$ and $\mathbf{p}$ can be collected into the Lorentz-covariant object $P^\mu$, the quantity $m$ behaves as a Lorentz scalar from the perspective of in-universe observers.

    In order to be able to cleanly substitute all of our quantities into Eq.~\eqref{Energy_conservation_with_functions_of_momenta}, we desire to be able to write $E=E(\vec{p})$. Noting that both $\vec{p}$ and $\gamma$ are functions of $\vec{v}$, we can perform some simple algebra to obtain the following familiar expression for $\gamma=\gamma(\vec{p})$:
    \begin{equation}
        \gamma(\vec{p}) = \sqrt{1+\dfrac{p^2}{m^2c^2_\sonic}};
    \end{equation}
    where $p^2=\vec{p}\cdot\vec{p}$. Substituting this expression for $\gamma$ into our expression for energy in Eq.~\eqref{Eq:Relativistic_particle}, we obtain $E=E(\vec{p})$. It proves more useful for our purposes to consider $E^2(\vec{p})$, however (the form of which should also be quite familiar):
    \begin{equation}
        E^2(\vec{p})
    =
        p^2c^2_\sonic+m^2c^4_\sonic
    \label{Eq:Internal_particle_dispersion_relation}.
    \end{equation}
    Now that we have $E=E(\vec{p})$, we can proceed precisely as we described in Sec.~\ref{Sec:Paper_Strategy}. Rather than performing all of our substitutions and then algebraically simplifying afterwards, we choose to perform the reverse process. Square the energy and momentum conservation equations [Eqs.~\ref{Energy_conservation_with_functions_of_momenta} and \eqref{Eq:Final_momentum_from_momentum_conservation}] to obtain the following:
    \begin{align}
        E^2_\final(\vec{p}_\final) &= \left\{E_\init(\vec{p}_\init) - \hbar\left[\omega_\final(\vec{k}_\final) - \omega_\init(\vec{k}_\init)\right]\right\}^2,
        \nonumber \\*
        &=
	E^2_\init
-
	2 \hbar
	E_\init
		\left(
		\omega_\final
		-
		\omega_\init
	\right)
+
	\hbar^2\left(
	\omega^2_\final
- 2\omega_\init
	\omega_\final
+
	\omega^2_\init
	\right)
	;
	\label{Eq:Energy_conservation_squared}
	\\
	p^2_\final &= \left[\vec{p}_\init - \hbar\left(\vec{k}_\final - \vec{k}_\init\right)\right]^2,
	\nonumber \\*
	&= p^2_\init - 2\hbar{p_\init}k_\final\cos\theta + 2\hbar{p_\init}{k_\init}
	\nonumber
	\\*
	&\qquad+ \hbar^2\left(k^2_\final + k^2_\init\right) - 2\hbar^2 k _\final k_\init\cos\theta
	\label{Eq:Momentum_conservation_squared}.
    \end{align}
    Note that the $\cos\theta$ terms in Eq.~\eqref{Eq:Momentum_conservation_squared} arise as a result of the experimental scenario that we are considering as per Fig.~\ref{Fig:Compton Scattering not-boosted and boosted}. In Eq.~\eqref{Eq:Energy_conservation_squared} we have suppressed the explicit functional dependency on momentum for notational brevity: this dependency still applies, of course.
    From here, we can directly substitute Eq.~\eqref{Eq:Internal_particle_dispersion_relation} into Eq.~\eqref{Eq:Energy_conservation_squared} wherever the square of the particle's energy appears. Doing so, and utilizing the square of the conservation of momentum [Eq.~\eqref{Eq:Momentum_conservation_squared}], we obtain the following expression:
  \begin{align}
&
	c_\sonic
	{p_\init}
	\hbar
	\omega_\init
-
	\hbar
	E_\init
	\omega_\init
	\nonumber
	\\*
&
\qquad
=
	\hbar
	\omega_\final
	\left(
		c_\sonic
		{p_\init}
	+
		\hbar
		\omega_\init
	\right)
	\cos \theta
-
	E_\init
	\hbar
	\omega_\final
-
	\hbar^2
	\omega_\init
	\omega_\final
	,
\end{align}
where we have made use of the fact that $\omega=c_\sonic\left\vert{\vec{k}}\right\vert$. From here, it is a simple matter to isolate $k_\final$. Doing so yields the following expression:
\begin{equation}
    k_\final
=
	\dfrac{\left(E_\init
		-
		c_\sonic p_\init\right)k_\init
		}{E_\init
		-
		c_\sonic p_\init
		+
		\left(1-\cos\theta\right)\left(\hbar\omega_\init + c_\sonic p_\init\right)
		}
	.
\label{Eq:InternalParticleKSolution}
\end{equation}
Some simple algebraic manipulations also allow us to phrase this in the following way:
\begin{align}
	\dfrac{
		\omega_\final}{
		\omega_\init}
& =
    \left[
        1
        +
        \left(
        \frac
            {\hbar\omega_\init + c_\sonic{p_\init}}
            {E_\init - c_\sonic {p_\init}}
        \right)
        (1-\cos \theta)
    \right]
    ^{-1}
	.
\label{Eq:PhononEnergyRatio}
\end{align}
Using the explicit forms of $E$ and $\vec{p}$ as given in Eq.~\eqref{Eq:Relativistic_particle}, and recalling that our experiment is initialized such that $\absvec{\beta_\init}=\beta_\init$ (with $\beta_\init>0$), we can further rewrite Eq.~\eqref{Eq:PhononEnergyRatio} entirely in terms of dimensionless quantities:
\begin{align}
	\frac{
		\omega_\final}{
		\omega_\init}
& =
    \left[
        1
        +
        \left(
		    \frac{\hbar\omega_\init}{\gamma_\init mc^2_\sonic}
		    +
		    \beta_\init
        \right)
        \left(
        \frac{1-\cos \theta}{1-\beta_\init}
        \right)
    \right]
    ^{-1}
	.
\label{Eq:PhononEnergyRatio_simplified}
\end{align}
This is just the kinematic relationship for ordinary Compton scattering~\cite{comptonQuantumTheoryScattering1923} as viewed from a boosted frame, except the boost is with respect to the speed of sound $c_\sonic$ rather than the speed of light. In other words, this is the kinematic description of a \textit{sonic analogue to Compton scattering}.

\subsubsection{The comoving in-universe observer frame kinematics of phonon scattering from internal particles}
\label{Subsec:InternalParticleComoving}
We now possess an unambiguous kinematic description of phonon scattering from an internal particle in the laboratory frame. Here we reexpress this from the perspective of an in-universe observer who is in the frame that is comoving with the particle prior to scattering.
We use Eqs.~\eqref{Eq:Transformation for initial frequency} and~\eqref{Eq:Transformation for final frequency}
	to write
\begin{align}
    \frac
        {\omega^\prime_\final}
        {\omega^\prime_\init}
&=
    \left(
    \frac
        {1+\beta_\init}
        {1+\beta_\init \cos \theta'}
    \right)
     \frac
        {\omega_\final}
        {\omega_\init}
    .
\end{align}
Plugging in Eq.~\eqref{Eq:PhononEnergyRatio_simplified} and using Eqs.~\eqref{Eq:Transformation for initial frequency}--\eqref{Eq:TransformationOneMinusCosTheta} to rewrite all quantities from the comoving frame, this becomes
\begin{align}
\frac
        {\omega^\prime_\final}
        {\omega^\prime_\init}
&=
    \left(
    \frac
        {1+\beta_\init}
        {1+\beta_\init \cos \theta'}
    \right)
\nonumber \\*
&\qquad \times
    \left[
        1
        +
        \left(
		    \frac{\hbar\omega^\prime_\init}{mc^2_\sonic}(1+\beta_\init)
		    +
		    \beta_\init
        \right)
        \left(
        \frac{1-\cos \theta'}{1+\beta_\init \cos \theta'}
        \right)
    \right]
    ^{-1}
    .
\end{align}
Straightforward algebra simplifies this to
\begin{align}
\frac
        {\omega^\prime_\final}
        {\omega^\prime_\init}
&=
    \left[
    		1
    		+
    		\left(\frac{\hbar\omega^\prime_\init}{mc^2_\sonic}\right)\left(1-\cos\theta^\prime\right)
    \right]^{-1}
    .
    \label{Eq:comoving_ratio_of_phonon_energies}
\end{align}
Notice that all dependence on $\beta_\init$ has disappeared in the final form, as it must since both types of particle respect the sonic Lorentz symmetry.

	While Eq.~\eqref{Eq:comoving_ratio_of_phonon_energies} may not be in its most familiar form, this is precisely the kinematic description of Compton scattering in the rest frame of a particle prior to scattering (with, of course, the understanding that all references to $c$ are replaced by $c_\sonic$). Recalling that our dispersion relation for phonons is taken to be linear, this expression can be easily cast into perhaps its most well known form:
	\begin{equation}
	    {\lambda^\prime_\final-\lambda^\prime_\init} = \dfrac{h}{mc_\sonic}\left(1-\cos\theta^\prime\right).
	    \label{Eq:Compton_formula}
	\end{equation}
	Note that the comoving in-universe observer frame's kinematic description of phonon scattering from internal particles makes no explicit reference to any velocity. This should not be too surprising, given the relativistic form of the energy-momentum relation that we chose for internal particles. It is important to note that we did not manually set $\beta_\init=0$ to obtain this result, though: all instances of $\beta_\init$ and $\gamma_\init$ were removed from the kinematic expression purely via the use of valid algebraic manipulations. The resulting expression, however, is entirely equivalent to taking Eq.~\eqref{Eq:PhononEnergyRatio_simplified}, setting $\beta_\init=0$ (and hence, $\gamma_\init=1$), and then appending primes to the remaining operationally determined quantities---a method that should be familiar to any student of special relativity.

	At this point, readers may find themselves wondering why we seem to be rederiving results that are well known in QFT~\cite{peskinIntroductionQuantumField2018a}. While the na\"{i}ve method of setting $\beta_\init=0$ in Eq.~\eqref{Eq:PhononEnergyRatio_simplified} does indeed provide a shortcut to the comoving in-universe description of scattering [i.e.,~without explicitly needing to consider the transformation equations Eqs.~\eqref{Eq:Transformation for initial frequency}--\eqref{Eq:TransformationOneMinusCosTheta}], this only works for scattering from particles that are sonically Lorentz obeying in nature. Specifically, the equivalence of these two methods is a result of the fact that the energy-momentum relation for internal particles transforms covariantly under the same sonic Lorentz transformation that applies to in-universe observer reference frames.\footnote{As noted in passing earlier, the internal particle's energy and momentum can be collected into the four-momentum $P^\mu$. With $P^\mu\coloneqq(E/c_\sonic,\vec{p})$, one can use Eq.~\eqref{Eq:Relativistic_particle} ($E=\gamma{mc^2_\sonic}$ and $\vec{p}=\gamma{m\vec{v}}$) to verify that $P^\mu$ is indeed a Lorentz-covariant object.} External particles, however, are \textit{not} sonically Lorentz obeying in nature---their energy-momentum relations do not transform covariantly under the sonic Lorentz transformation---and so the shortcut that happened to work with internal particles \textit{does not} apply when we are considering external particles.

	Reobtaining the familiar expression for Compton scattering by utilizing the simple approach that we first detailed in Sec.~\ref{Sec:Paper_Strategy} provides a sanity check for our approach, and it provides a familiar example through which to elucidate some important details regarding the derivation.

\subsection{Phonon scattering from external particles}
\label{SubSec:ExternalParticleKinematic}

\subsubsection{The laboratory frame kinematics of phonon scattering from external particles}
\label{SubSec:ExternaParticleScatteringLab}

As we have stated before, external particles are particles that are \textit{not} sonically Lorentz obeying. That is to say, the dynamical description of external particles is not covariant with respect to the sonic Lorentz symmetry of the analogue-gravity medium. This, of course, leaves a wide range of possibilities in terms of selecting energy-momentum relations for external particles. For our purposes here, we choose specifically to consider our external particle to be an ordinary quantum-mechanical particle with the usual Newtonian energy-momentum relation:
\begin{align}
    E &= \dfrac{p^2}{2m} & \vec{p}&=m\vec{v} \label{Eq:Newtonian particle}.
\end{align}
Following the approach described in Sec.~\ref{Sec:Paper_Strategy}, we directly substitute our expression for energy into Eq.~\eqref{Energy_conservation_with_functions_of_momenta} and utilize the conservation of momentum [Eq.~\eqref{Eq:Final_momentum_from_momentum_conservation}] to eliminate any reference to $\vec{k}_\final$. This leads to an equation that is quadratic in $k_\final$:
\begin{align}
    \frac{
    \hbar^2
        k_\final^2
        }{
        2m}
    -
        \hbar k_\final
        \left[
            \frac{
            \left(\hbar{k_\init}+{p_\init}\right)\cos\theta
            }{
            m}
        -
            c_\sonic\right]
            &
            \nonumber\\
    +
    \hbar
    k_\init
    \left(
        \frac{
            \hbar k_\init
        +
            2 p_\init
            }{2m}
        -
            c_\sonic
        \right)
        &
    =
    0.
    \label{Eq:ConserEnergyExternal1}
\end{align}
In the fully sonically Lorentz obeying case (that is, phonon scattering from internal particles) the conservation of energy and momentum lead to a linear equation in $k_\final$, and thus for a given set of initial experimental parameters and for a given scattering angle $\theta$ there existed only one possible value of $k_\final$. In this case, however, we have a quadratic equation, and so for a given set of initial experimental parameters and for a fixed scattering angle of $\theta$, there are two possible values that $k_\final$ can take. We can solve Eq.~\eqref{Eq:ConserEnergyExternal1} for $\hbar k_\final$ and multiply the solutions by $c_\sonic$ to obtain solutions of the following form for the final energy of the phonon:
\begin{align}
	c_\sonic \hbar k_{\final,(1,2)}
=
\hbar\omega_{\final,(1,2)}
=
	c_\sonic
	\bar{B}
\pm
	c_\sonic
	\sqrt{
		\bar{B}^2
	-
		\bar{C}}
	.
\label{Eq:FinalPhononEnergy}
\end{align}
where $c_\sonic \hbar  k_{\final,(1)}  $ (or equivalently, $\hbar\omega_{\final,(1)}$) is taken to be the solution with the positive sign, $c_\sonic \hbar k_{\final,(2)}$ (or equivalently, $\hbar\omega_{\final,(2)}$) is taken to be the solution with the negative sign, and%
\begin{subequations}
\label{eqs:BCbar}
\begin{align}
	\bar{B}
&
=
	(\hbar{k_\init}
+
	 {p_\init})
	 \cos \theta
-
	m c_\sonic
	,
	\\*
	\bar{C}
&
=
	\hbar{k_\init}
	\left(
		\hbar{k_\init}
	+
		2{p_\init}
	-
		2 m c_\sonic
	\right)
	.
\end{align}
\end{subequations}
Note that the bar ($\bar{~}$) serves only as a notational label; it has no particular physical or mathematical meaning. The quantity $\bar{B}$ has units of a three-momentum, whereas the quantity $\bar{C}$ has units of the square of a three-momentum.

The term $c_\sonic\bar{B}$ in Eq.~\eqref{Eq:FinalPhononEnergy} expands to produce terms of the form $c_\sonic \hbar{k_\init}$, $mc^2_\sonic$, and $c_\sonic {p_\init}$. The first of these three terms is precisely the initial energy of the phonon,~$\hbar \omega_\init$. The remaining two terms have the structure of relativistic terms: by analogy to actual relativity, the second term appears to have the structure of a sonic analogue to rest mass energy, and the third term has the structure of the initial (relativistic) kinetic energy of the particle. With that said, these final two terms are \textit{not} actually descriptions of the external particle's energy because the external particle is \textit{not} sonically relativistic. Nonetheless, it is interesting to note that terms of a relativistic form appear naturally in the description of our nonrelativistic external particle.

As for the case of phonon scattering from internal particles, we also express the kinematic description of phonon scattering from external particles in terms of the ratio $\omega_\final/\omega_\init$:
\begin{align}
		\frac{
		\omega_{\final,(1,2)}}
		{
		\omega_\init
		}
=
	B
\pm
	\sqrt{
		B^2
		-
		C}
		,
\label{Eq:EnergyRatioPhononExternal}
\end{align}
where, as per above, $\omega_{\final,(1)}$ is taken to be the positive solution and $\omega_{\final,(2)}$ is taken to be the negative solution. The quantities $B$ and $C$ are related to their barred versions and are given as follows:
\begin{subequations}
\label{eqs:BC}
\begin{align}
	B
&
=
	\frac{c_\sonic \bar{B}}{\hbar \omega_\init}
=
	\left(
		1
	+
		\frac{
			c_\sonic
			{p_\init}}
			{\hbar \omega_\init}
	\right)
	\cos \theta
-
	\frac{
		m c^2_\sonic}
		{\hbar
		\omega_\init}
	;
	\\
    C
&
=
	\frac{c^2_\sonic \bar{C}}{(\hbar \omega_\init)^2}
=
	1
+
	\frac{2
			c_\sonic
			{p_\init}}
			{\hbar \omega_\init}
-
	\frac{2
		m
		c^2_\sonic}
		{\hbar
		\omega_\init}
	.
\label{Eq:InitialToFinalEnergyRatioExternalSimplified}
\end{align}
\end{subequations}
The expressions $B$ and $C$ will prove useful when we move into the comoving in-universe observer reference frame. Before proceeding to determine the description of kinematics from that frame, we shall take a moment to address the nature of the two solutions for the kinematic description of phonon scattering from external particles.

\subsubsection{Physical meaning of the two solutions}

We will now focus on understanding the physical meaning of the two solutions given by Eq.~\eqref{Eq:FinalPhononEnergy}. Scattering from either type of particle should always yield a real and positive value of the ratio of the final-to-initial phonon energies. In the case of scattering from the external particle, the requirement that the ratio of phonon energies be real puts restrictions on the allowed values of $\theta$ for given initial ${p_\init}$ and $\hbar \omega_\init = c_\sonic \hbar{k_\init}$ since the discriminant of the square-root in Eq.~\eqref{Eq:FinalPhononEnergy} must be non-negative. From Eq.~\eqref{Eq:EnergyRatioPhononExternal} we see that this leads to two conditions: %

\paragraph*{Condition 1.---} \hspace{-2mm}Solutions 1 and 2 are \textit{both} guaranteed to be real and positive when the following conditions are both satisfied:
\begin{align}
	\bar C
&\ge
    0
	&
	& \text{and}
	&
    \bar B
&\ge
	\sqrt{\bar C}
    .
\label{Eq:Condition1}
\end{align}
This condition directly translates into an inequality in $\cos \theta$, restricting the scattering angle range as follows:
\begin{align}
	1
\ge
	\cos \theta
\ge
	\frac{
		mc_s
	+
		\sqrt{
			\hbar{k_\init}
			\left(
				\hbar{k_\init}
			+
				 2{p_\init}
			-
				2 m c_s
			\right)
		}
	}{
	(\hbar{k_\init} + {p_\init})
	}
	,
\end{align}
with
\begin{align}
	\hbar{k_\init}
	\left(
		\hbar{k_\init}
	+
		2{p_\init}
	-
		2 m c_\sonic
	\right)
	\ge
	0
	\quad
	\Longrightarrow
	\quad
	m c_\sonic
	\le
	\frac{
		\hbar{k_\init}
	+
		2
		{p_\init}
	}{2}
	.
\end{align}
\paragraph*{Condition 2.---}\hspace{-4mm} If $\bar C \le 0$ then solution 1 is always real and positive (note that $\bar B^2$ is always positive). This means that there aren't any restrictions on the scattering angle, and so $\cos\theta$ can take the usual range of values:
\begin{align}
	1
\ge
	\cos \theta
\ge
	-1
	.
\end{align}
Restrictions are however placed on the absolute values of the three-momenta $\hbar{k_\init}$ and ${p_\init}$ such that
\begin{align}
	\hbar{k_\init}
	\left(
		\hbar{k_\init}
	+
		2{p_\init}
	-
		2 m c_\sonic
	\right)
	\le
	0
	\quad
	\Longrightarrow
	\quad
	m c_\sonic
	\ge
	\frac{
		\hbar{k_\init}
	+
		2
		{p_\init}
	}{2}
	.
\end{align}
These conditions will play a very important role in the calculation of the scattering cross section in Sec.~\ref{Sec:Scattering cross sections}.

\subsubsection{The comoving in-universe observer frame kinematics of phonon scattering from external particles}

As we have discussed previously, the kinematic description of phonon scattering from external particles in the comoving in-universe observer frame is \textit{not} simply given by taking the laboratory frame description and setting ${\beta_\init}=0$. The observers' geometric measurements constitute a Lorentz-covariant object ($X^\mu$), but the energy and momentum of the external particle \textit{cannot} be collected into a sonically Lorentz-covariant object. As a result of this mismatch in symmetry groups between the dynamically relevant physical quantities (energy and momentum) and the reference frames of observers, the kinematic description of phonon scattering from external particles is \textit{not} Lorentz covariant. We take then the approach that we discussed in Sec.~\ref{Sec:Paper_Strategy} (and which we previously applied to the kinematics for phonon scattering from internal particles in Sec.~\ref{Subsec:InternalParticleComoving}) and use the relations given by Eqs.~\eqref{Eq:Transformation for initial frequency}--\eqref{Eq:TransformationOneMinusCosTheta} to replace the value of all laboratory frame quantities with their comoving in-universe observer frame values.

Substituting these formulas into Eq.~\eqref{Eq:FinalPhononEnergy}, some basic algebraic manipulations lead to the following description of phonon scattering from external particles in the comoving in-universe observer frame:
\begin{align}
	c_\sonic
	\hbar k'_{\final, (1,2)}
&
=
	\hbar
	\omega'_{\final,(1,2)}
	\nonumber
	\\*
&
=
    \frac{c_\sonic}{
        \gamma_\init
		\left(
			1
		+
			\beta_\init
			\cos \theta'
		\right)}
	\left(
	\bar{B'}
\pm
	\sqrt{
		\bar{B}'^2
	-
		\bar{C}'}
	\right)
	,
\end{align}
where%
\begin{subequations}
\label{eqs:BCbarprime}
\begin{align}
	\bar{B}'
&
=
			\left(
				D \hbar{k^\prime_\init}
			+
				 m c_\sonic
				{\beta_\init}
			\right)
			\left(
				\frac{
					\cos \theta'
				+
					\beta_\init
					}{
					1
				+
					\beta_\init \cos \theta'}
			\right)
		-
			m c_\sonic
	,
	\\*
	\bar{C}'
&
=
		D \hbar{k^\prime_\init}
		\left(
			D\hbar{k^\prime_\init}
		+
			2 m c_\sonic {\beta_\init}
		-
			2 m c_\sonic
		\right)
	.
\end{align}
\end{subequations}
$\bar B'$ and $\bar C'$ are just $\bar B$ and $\bar C$, respectively, written in the comoving ($'$) coordinates using
Eqs.~\eqref{Eq:Transformation for initial frequency}--\eqref{Eq:TransformationOneMinusCosTheta}.
Note that we have used the Doppler factor~$D$ as defined in Eq.~\eqref{Eq:Doppler_factor}, with $\beta=\beta_\init$, for notational brevity in these expressions.

In the in-universe observer frame, the dimensionless ratio of the final to initial frequencies of the phonon can be given by the following relation:
\begin{align}
	\dfrac{
		\omega'_{\final, (1,2)}
		}{
		\omega'_\init}
=
    \frac{
        (1+\beta_\init)
        }
        {
        (1+\beta_\init
        \cos \theta')}
    \left(
        B'
    \pm
        \sqrt{
            B'^2
        -
            C'}
    \right)
	,
\label{Eq:Solution to comoving frame xi in terms of boosted parameters classical}
\end{align}
where here $B'$ and $C'$ are
\begin{subequations}
\label{eqs:BCprime}
\begin{align}
	B'
&
=
	\frac{
				\cos \theta'
			+
				\beta_\init
		}{
			1
		+
			\beta_\init
			\cos \theta'
		}
	-
	\frac{
			m c^2_\sonic
			\left(
				1
			-
				\beta^2_\init
			\right)
		}{
		D
		\hbar
		\omega'_\init
		\left(
			1
		+
			\beta_\init
			\cos \theta'
		\right)
		},
	\\*
    C'
&
=
    1
    -
	\frac{
			2
			m
			c^2_\sonic
		}
		{
		D
		\hbar
		\omega'_\init
		}
		(1-\beta_\init)
	.
\end{align}
\end{subequations}
In fact, the reader can verify, using Eqs.~\eqref{Eq:Transformation for initial frequency}--\eqref{Eq:TransformationOneMinusCosTheta}, that $B'$ and $C'$ are just $B$ and $C$ expressed in comoving ($'$) coordinates. The kinematic description of phonon scattering from an internal particle and the comoving in-universe observer frame [Eq.~\eqref{Eq:comoving_ratio_of_phonon_energies}] made no reference to the velocity of the particle (and hence the comoving observer), whereas the kinematic description of phonon scattering from an external particle [Eq.~\eqref{Eq:Solution to comoving frame xi in terms of boosted parameters classical}] does retain reference to the velocity of the particle (and hence the comoving observer). The presence of velocity dependency in the comoving in-universe observer's description of phonon scattering from external particles means that, in principle, an in-universe observer could use such scattering experiments to identify their state of motion with respect to their analogue universe. We shall come back to this point in Sec.~\ref{Subsec:Using Lorentz-violating sonic Compton scattering to determine absolute motion}.

The forbidden scattering angles and three-momenta that arise in the laboratory frame description of phonon scattering from external particles take on the following forms in the comoving in-universe observer frame:

\paragraph*{Condition 1.---} \hspace{-2mm}Solutions 1 and 2 are \textit{both} guaranteed to be real and positive when the following conditions are both satisfied:
\begin{align}
	\bar C'
&\ge
	0
	&
	& \text{and}
	&
	\bar B'
&\ge
	\sqrt{\bar C'}.
\label{Eq:RealPositiveConditionsInUniverse}
\end{align}
This condition translates into the following restrictions on the scattering angle $\cos \theta'$:
\begin{align}
&
	1
\ge
	 \cos \theta'
\ge
    \frac{
        mc_\sonic
        \left(
            1
        -
            \beta^2_\init
        \right)
        -
            \beta_\init D \hbar k'_\init
        +
            \sqrt{\bar C'}
        }{
            D \hbar k'_\init
        -
            \beta_\init \sqrt{\bar C'}
        },
        \label{Eq:Comoving_angular_restrictions}
\end{align}
with
\begin{align}
	mc^2_s
\le
	\frac{
		D
		\hbar
		\omega'_\init
		}{2 (1-\beta_\init)}
		.
\end{align}

\paragraph*{Condition 2.---}\hspace{-4mm} If $\bar C' \leq 0$ then solution 1 is always real and positive [note that $(\bar B')^2$ is always positive]. In this case, there are no restrictions on the scattering angles, but the absolute values of the three-momenta are constrained:
\begin{align}
	1
\ge
	\cos \theta'
\ge
	-1
	,
\end{align}
with
\begin{align}
	mc^2_s
\ge
	\frac{
		D
		\hbar
		\omega'_\init
		}{2 (1-\beta_\init)}
		.
\end{align}

\section{Scattering cross sections}
\label{Sec:Scattering cross sections}

\renewcommand{\op}[1]{\hat{#1}}

We now have an understanding of the kinematics of scattering for both the case of a phonon scattering from an internal particle (obeying sonic Lorentz symmetry), and the case of a phonon scattering from an external particle (not obeying sonic Lorentz symmetry). Our goal now is to understand whether, by analyzing the cross sections for phonons scattering on external particles, in-universe observers can infer anything about their velocity with respect to the laboratory reference frame.

\begin{figure}[tb]
	\includegraphics
	[width=\linewidth,trim=0 1.5cm 0 1.5cm,clip]
	{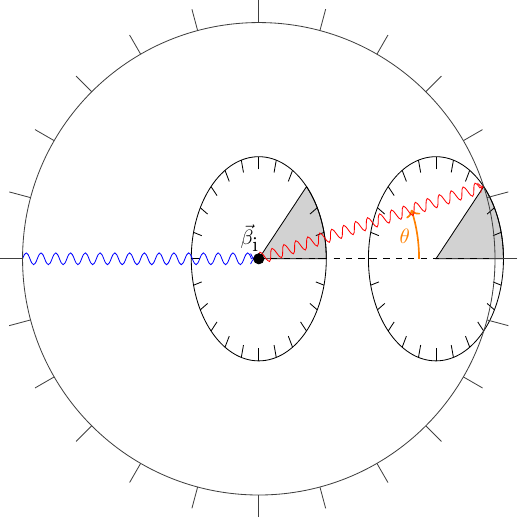}
	\caption{Phonon scattering from an external particle as viewed in the laboratory frame. The large circle centered on the scattering event (clipped at the top and bottom) represents a 2D slice of a spherical array of detectors that is stationary with respect to the laboratory frame. The two ellipses represent 2D slices of a \textit{single array of detectors} (comoving with the particle prior to scattering) at two different moments in time. Ticks on the circumference of both detector arrays are separated by $15^\circ$ \textit{in the frame of that detector array}.
	The scattering event occurs at the moment in time for which the geometric centers of both detector arrays are coincident. At some later point in time, the scattered phonon is again coincident with both detector arrays: the laboratory frame detector array detects the phonon at the detector located at an angle $\theta$ above the $z$ axis (here $\sim17.8^\circ$), which in the comoving frame corresponds to the detector located at the angle $\theta^\prime$ above the $z^\prime$ axis (here $45^\circ$, indicated by the shaded wedge in the moving detector array). In this particular example $\beta_\init=0.75$, corresponding to $\gamma_\init=4/\sqrt{7}\approx{1.5}$ (which is the factor by which the moving detector array is contracted in its direction of motion). Fig~\ref{Fig:Compton Scattering not-boosted and boosted}(a) shows the same scattering event indicated here, whereas Figure~\ref{Fig:Compton Scattering not-boosted and boosted}(b) shows this scattering event in the frame of the moving detector array.}
	\label{Fig:Compton Scattering cross sections}
\end{figure}

In pursuit of this, we need to devise some appropriate quantum toy model for sonic particle scattering that captures the important features of actual particle scattering, such as Compton scattering~\cite{comptonQuantumTheoryScattering1923,peskinIntroductionQuantumField2018a}. The physics in the laboratory frame is unambiguous when we have such a model, and the method by which we determine the comoving in-universe observer description of events is by transforming observable quantities appropriately. We also continue to define all measurements operationally: we imagine that in-universe observers take an operational approach to detecting scattered phonons by using arrays of detectors that are, in their own frame, seen to be spherical and seen to be centered on the scattering event. Figure~\ref{Fig:Compton Scattering cross sections} highlights what this looks like in the laboratory frame. By performing several scattering events to create a statistical distribution and counting how many times any single detector clicks, we can obtain the \textit{reaction rate} $\mathcal R$---i.e.,~the number of scattering events per unit of volume per unit of time in a particular direction---to obtain the differential cross section.

In the following, we will start by carefully deriving the scattering cross section equations for external particles.
Our system is a ``hybrid'' one, composed of three parts: the reference frames that we use to operationally perform measurements, the phonon, and the external particles. Only two of the three---the phonon and the reference frames---are always fully sonically relativistic, hence we cannot simply apply the QFT definition of the cross section
since Lorentz covariance is deeply embedded in the derivation~\cite{peskinIntroductionQuantumField2018a}. On the other hand, we have to upgrade the ordinary (nonrelativistic) quantum-mechanical scattering theory as this is usually derived only in the laboratory reference frame~\cite{Sakurai:1167961}, while we want to be able to transform between Lorentz-obeying reference frames.\footnote{The operational method of using arrays of detectors for the derivation of scattering cross section is a useful tool in our hybrid system, %
as we can easily transform the position of the detectors (with the transformation rules derived in Sec.~\ref{Sec:PhononScattering}) and use them to easily redefine a reaction
rate in the new reference frame.}

\subsection{General cross section definition}

In quantum physics, the cross section measures the transition probability for a specific process to happen in the scattering of two or more particles~\cite{cannoniLorentzInvariantRelative2017}. The total cross section is defined as
\begin{align}
	\sigma
=
	\frac{
		\mathcal R
		}{
		\Phi}
	,
\label{Eq:CrossSectionGeneralDefinition}
\end{align}
where $\mathcal R$ is the \emph{reaction rate} (transition probability per unit time per unit volume), and $\Phi$ is the incident flux of particles. The cross section~$\sigma$ has units of area and qualitatively represents the effective cross sectional area (hence the name) that the particle presents to the incoming beam of particles to be scattered, and it will depend on the momentum of the incoming particles.

The cross section as defined in Eq.~\eqref{Eq:CrossSectionGeneralDefinition}
can be expressed more explicitly as
\begin{align}
	\sigma[\mathcal{S}|\init]
&
=
	\frac{1}{\Phi}
	\sum_{\final\in \mathcal{S}}
	\frac{w_{\init\rightarrow \final}}{L^3}
=
	\frac{1}{\Phi}
	\sum_{\vec k_\final}
	\sum_{\vec p_\final}
	\frac{w_{\init\rightarrow \final}}{L^3}
	,
\label{Eq:CrossSectionDetailedDefinition}
\end{align}
where $\ket \init$ and $\ket \final$ represent the initial and final states, respectively (often shown with ket symbols omitted), and where $\mathcal{S}$ is the subset of all final states that are different from~$\init$ (in order to exclude the case of no scattering). We use box normalization (volume~$L^3$) for our particles, so that the reaction rate~$\mathcal R$, which is a transition rate per unit volume, becomes $w_{\init\rightarrow \final}/L^3$, with
\begin{align}
	w_{\init \rightarrow \final}
=
	\frac {2 \pi} {\hbar}
	\abss{ \bra \final \op H_\text{int}
	\ket \init }^2
	\delta(E_{\final \init})
	,
\label{Eq:TransitionRate}
\end{align}
being the transition rate from an initial state $\ket \init$ at $t = -\infty$ to a final state $\ket \final$ at time $t = T$, given some interaction Hamiltonian $\op H_\text{int}$. The symbol $E_{\final \init}$ represents the change in total energy of the system (as measured in the lab), and the states and the interaction Hamiltonian are all defined in a Schr\"odinger picture. For more details on how to derive Eq.~\eqref{Eq:TransitionRate}, please see Ref.~\cite{Sakurai:1167961}.

The notation $\sigma[\mathcal{S}|\init]$ represents the cross section for scattering to any final state $\ket \final \in \mathcal{S}$ (different from the initial one) given the particular initial state~$\ket \init$. The last form of Eq.~\eqref{Eq:CrossSectionDetailedDefinition} is the one we will use for our calculation.

\subsection{Cross section in the in-universe observers' reference frame}
\label{subsec:CrossSectionTransformation}

Before going into the details of the cross section derivation, we want to clarify the behavior of the cross section $\sigma$ when viewed from the frame of an in-universe observer in motion with respect to the laboratory frame.

Since our model lacks full (sonic) Lorentz invariance, we must reexamine some of the basic assumptions about cross sections---most especially, how it behaves when viewed from a moving frame. In ordinary QFT, which has full Lorentz invariance, the cross section behaves exactly like a cross sectional area: it is invariant under boosts in the direction of motion of the incoming particle to be scattered (assuming a stationary target particle) but not generally in other directions~\cite{peskinIntroductionQuantumField2018a}. We cannot rely on this being true in our model. To examine how the cross section behaves in this hybrid model in which some pieces obey sonic Lorentz symmetry, we return to its basic form, Eq.~\eqref{Eq:CrossSectionGeneralDefinition}, which is a ratio of the reaction rate~$\mathcal R$ to the incoming flux~$\Phi$. We will examine how each of these would be perceived by an in-universe observer comoving with the external particle.

\subsubsection{Reaction rate}

The reaction rate~$\mathcal R$ is a transition rate per unit volume. This can be reexpressed as the total transition \textit{probability} per unit \textit{spacetime volume}.

A spacetime volume element is invariant under a Lorentz transformation, and thus it remains invariant when measured by in-universe observers. Thus, the denominator of~$\mathcal R$ is invariant.

Now let us examine the numerator. The probability of some particular event happening also remains invariant, even though the different observers may disagree on the particular descriptors used to specify the event (e.g., direction, duration, etc.). If a given detector clicks, then all observers will agree that the detector clicked. At the end of a given set of experiments, all observers---regardless of state of motion---will therefore agree on the list of detectors that clicked, and how many times each detector clicked. Thus, the probability of a physical detector clicking must also be independent of an observer's state of motion.

This means that the transition probability for any given given event~$\init \to \final$ must be invariant. A moving observer would interpret this as the probability for the transition~$\init' \to \final'$, where $\init'$ and $\final'$ are the initial and final states as interpreted by the moving observer, but the numerical value of this probability would be the same for both observers. The total transition probability is just the sum of all relevant individual transition probabilities of this form. Since this sum includes all possible final states (except the initial one), all observers must agree on this value. Thus the total transition probability (numerator of~$\mathcal R$) is invariant.

Combining the above two results, we see that the reaction rate~$\mathcal R$ is itself invariant, being the ratio of two invariant quantities: total transition probability and amount of spacetime volume.

\subsubsection{Flux}

The behavior of the flux~$\Phi$ under a Lorentz transformation is the subject of much discussion, even in a fully relativistic theory such as QFT. A recent article~\cite{cannoniLorentzInvariantRelative2017} explains clearly the subtleties of this topic and provides the definitive resolution. The key results of that work also apply to our setup, and we repeat them here, along with a discussion of how our particular setup modifies them (or not).

The approach will be to write the flux in a manifestly Lorentz-invariant form based on the sonic analogue to the four-current, which we denote as
\begin{align}
    J^\mu
&\coloneqq
\begin{pmatrix}
	n c_\sonic \\
	n \vec v
	\end{pmatrix}
	,
\end{align}
with $n$ being the number density of particles (in some frame) and $\vec v$ the three-velocity of the particle (as measured from the same frame). $J^\mu$ has units of speed times number density (or speed per unit volume), which has the intuitive interpretation of number of particles passing through a unit of area per unit of time.
$J^\mu$ is a sonic Lorentz-covariant four-vector and will transform as such when changing from one in-universe observer's frame to that of another. Crucially, notice that this is a kinematical quantity, not a dynamical one. It is based on a description of the motion of the particle and makes no reference to its particular dispersion relation.

The crucial observation from Ref.~\cite{cannoniLorentzInvariantRelative2017} is that the flux for a two-particle scattering experiment can be written in an invariant form using the four-currents~$J^\mu_1$ and $J^\mu_2$ for the two particles:%
\footnote{Reference~\cite{cannoniLorentzInvariantRelative2017} uses natural units. We have included the prefactor $c_\sonic^{-1}$ to get the units right in our case.}
\begin{align}
 	\Phi
=
    c_\sonic^{-1}
	\sqrt{
		(J_1 \cdot J_2 )^2
	-
		( J_1)^2
		( J_2)^2
	}
	,
\label{eq:FluxCurrents}
\end{align}
where 1 and 2 are the two scattering particles, the dot ($\cdot$) is the four-vector dot product (in our case, using $c_\sonic$ instead of~$c$), and $(J)^2 = J \cdot J$. Importantly, every in-universe observer will agree on the numerical value of this quantity, so we only need to calculate it in one frame: the laboratory frame.

In the laboratory frame, the four-currents for the phonon and the external particle, respectively, are
\begin{align}
    J^\mu_\sonic
&=
    n_\sonic
    \begin{pmatrix}
        c_\sonic \\ 0 \\ 0 \\ c_\sonic
    \end{pmatrix}
    ,
&
    J^\mu_\particle
&=
    n_\particle
    \begin{pmatrix}
        c_\sonic \\ 0 \\ 0 \\ v_\init
    \end{pmatrix}
    ,
\end{align}
where $n_\sonic$ and $n_\particle$ are the initial number density of the phonon and of the particle. Then,
\begin{align}
    J_\sonic \cdot J_\particle
&=
    n_\sonic n_\particle c_\sonic^2
    (1-\beta_\init)
    ,
\\
    (J_\sonic)^2
&=
    0
    ,
\\
    (J_\particle)^2
&=
    n_\particle^2 c_\sonic^2 (1-\beta_\init^2)
    .
\end{align}
Plugging these into Eq.~\eqref{eq:FluxCurrents} gives the formula for the flux in the laboratory frame:
\begin{align}
	\Phi
=
	n_\sonic
	n_\particle
	\left|
	c_\sonic
-
	v_\init
	\right|
	,
\end{align}
For our choice of normalization,%
\footnote{\label{foot:norm} There is much discussion about choosing a Lorentz-invariant normalization in QFT textbooks (e.g., Ref.~\cite{peskinIntroductionQuantumField2018a}), but this is merely a convention that has some calculational and conceptual utility in a fully relativistic setting. Importantly, it is not required for obtaining physically valid results---even in a fully relativistic setting. Any normalization will do as long as it is treated consistently throughout the calculation. And that is what we do.}
the number densities are $n_\particle=1/L^3$ and $n_\sonic = 1/L^3$, so the flux in the laboratory frame is
\begin{align}
    \Phi
&
=
    \frac{
        \left|
    	c_\sonic
    -
	    v_\init
	    \right|
	    }{
	    L^6}
=
    c_\sonic
    \frac{
        \left|
    	1
    -
	    \beta_\init
	    \right|
	    }{
	    L^6}
    .
\end{align}
This is the value of~$\Phi$ in all frames.

Nevertheless, a quick confirmation shows that this is indeed what in-universe observers comoving with the particle would calculate. Simply apply the appropriate Lorentz transformation to the currents, giving $J^{\mu\prime}_\sonic = D^{-1} n_\sonic (c_\sonic, 0, 0, c_\sonic)^\tp$ and ${J^\mu_\particle}' = \gamma_\init^{-1} n_\particle (c_\sonic, 0, 0, 0)^\tp$, and plug into Eq.~\eqref{eq:FluxCurrents}. The result for~$\Phi$ is unchanged.

\subsubsection{Total cross section}

The total cross section~$\sigma$ is the ratio of two sonic Lorentz scalars: $\mathcal R$ and $\Phi$. The numerical value of each of these quantities is therefore agreed upon by all in-universe observers, and so, therefore, is their ratio~$\sigma = \mathcal R/\Phi$. The total cross section is a sonic Lorentz scalar with respect to all in-universe observer frames moving in the $z$~direction.

Importantly, however, note that we have taken care to only assert that these quantities are invariant with respect to different states of motion of the \textit{observers}. The fact that the external particle violates the sonic Lorentz symmetry of the system means that the quantities in question are not necessarily invariant with respect to simultaneous boosts of all objects \textit{with respect to the medium}. In fact, as we will discover in the coming pages, experiments that appear to in-universe observers to have the same initial conditions---assuming the observers in each case are comoving with the particle---will nevertheless lead to different outcomes depending on their initial velocity with respect to the medium. In the language of Lorentz-violating extensions to the Standard Model~\cite{colladayLorentzviolatingExtensionStandard1998}, the cross section is invariant with respect to \textit{observer boosts} but not necessarily so with respect to \textit{particle boosts}.

\subsubsection{Differential cross section}

While we have shown above that the numerical value of the total cross section~$\sigma$ is agreed upon by all observers, we are actually interested in the \textit{differential} cross section~$d\sigma / d \cos \theta$ with respect to the scattering angle~$\theta$ in the laboratory frame and $d\sigma / d \cos \theta'$ with respect to scattering angle~$\theta'$ in the comoving frame.

We can illustrate this situation as follows. We imagine a large spherical array of detectors that is stationary with respect to the laboratory frame and centered on a phonon scattering event, as illustrated in Fig.~\ref{Fig:Compton Scattering cross sections}. Each detector subtends a small solid angle~$d\Omega(\theta, \phi)$ at a given orientation~$(\theta, \phi)$ as measured in the laboratory frame. Whether a particular detector has clicked or not is manifestly invariant, as is the probability of any physical detector clicking, as discussed above. So too would this reasoning apply to a spherical array of detectors comoving with the particle, as illustrated in Fig.~\ref{Fig:Compton Scattering cross sections}. Each of these detectors would subtend a solid angle $d\Omega'(\theta', \phi')$ at a given orientation~$(\theta', \phi')$ as measured in the comoving frame.

The total cross section can be split up into infinitesimal pieces~$d\sigma$, corresponding to scattering (in the laboratory frame) at angle~$\theta$ into a narrow ring of detectors for all azimuthal angles~$\phi$. We can reassemble these pieces into the total cross section:
\begin{align}
    \sigma = \int d\sigma = \int d\cos \theta\,
    \frac{d\sigma[\theta|\init]}{d\cos \theta}
    ,
\end{align}
where we make explicit the final scattering angle~$\theta$ and initial state~$\init$. Since the probability that some detector in that ring clicks is agreed upon by all observers, it is merely a question of kinematics as to how that ring appears to in-universe observers in a different state of motion. To find this, we first find $\theta'$ and $\init'$ using Eqs.~\eqref{Eq:TransformationCosTheta} and~\eqref{Eq:Transformation for initial frequency}, and then we note that
\begin{align}
    d\sigma[\theta|\init] = d\sigma[\theta'|\init']
    \label{Eq:DiffCrossSectionEquality}
\end{align}
by the scalar nature of~$\sigma$ and the argument above about the invariance of probabilities.

The comoving frame description of the scattering event is therefore obtained solely by appropriately transforming the laboratory-frame kinematic quantities into the comoving frame of the particle. From Eq.~\eqref{Eq:DiffCrossSectionEquality} we get the differential cross section in the comoving reference frame from the laboratory one~\cite{Tokimoto:780468}:
\begin{align}
	\frac{
		d \sigma[\theta'|\init']
		}{
		d \cos \theta'
		}
=
	\frac{
		d \sigma[\theta|\init]
		}{
		d \cos \theta
		}
	J[\theta,\theta']
	,
\label{Eq:DiffCSGeneralTransformation}
\end{align}
where $J[\theta,\theta']$ is the Jacobian of the coordinate transformations between reference frames,
\begin{align}
	J[\theta,\theta']
&
=
	\frac{d \cos \theta}{d \cos \theta'}
=
    \frac
	{1-\beta_\init^2}
    {(1 + \beta_\init \cos \theta')^2}
	 ,
\end{align}
obtained using the transformation Eq.~\eqref{Eq:TransformationCosTheta}. With these tools in hand, our approach is to calculate the differential cross section in the laboratory frame (since that is the frame in which we know the dynamics) and then use Eq.~\eqref{Eq:DiffCSGeneralTransformation} to express it from the comoving observers' perspective.

\subsection{Phonons, quantized external particles, and the interaction Hamiltonian in the laboratory frame}
\label{Subsec:Phonons, Quantized external particles, and the interaction Hamiltonian in the laboratory frame.}

In the laboratory frame, we assume that phonons are excitations of the analogue-gravity medium; for simplicity, we treat the phonon field as a scalar field.%
\footnote{Treating phonons as a scalar field is appropriate in systems that are isotropic. For systems that exhibit anisotropies, e.g.,~any system with a crystal structure, a vector field description must be used to fully capture all appropriate degrees of freedom~\cite{vonsovskiiPhononSpin1961}.}

We use standard quantum mechanics to describe external particles, as this suffices to capture all of their relevant quantum mechanical degrees of freedom. Note that the only degree of freedom that we endowed our external particle with (see Sec.~\ref{SubSec:ExternalParticleKinematic}) is its center-of-mass energy/momentum---we have not endowed the external particle with any additional degrees of freedom such as angular momentum.

So in summary, our recipe for a quantum-mechanical interaction Hamiltonian includes the following:
\begin{enumerate}
	\item First-quantized matter that describes the external particle.
	\item A single excitation of a second-quantized phonon field that describes the scattering phonon. %
	\item An interaction Hamiltonian depending on the phonon's field amplitude at the (quantized) position of the external particle.
\end{enumerate}
We define the phonon field as
	\begin{equation}
	    \op{\phi}(\vec{x})
	=
	    \frac{1}{L^{3/2}} \sum_{\vec k}
	   \sqrt{
	   \frac
	        {\hbar c_\sonic}
	        {2 k}
	   }
	   \left(
	        \op{a}_{\vec k}
	        e^{i \vec k\cdot \vec x}
	       +
	       	\op{a}^\dagger_{\vec k}
	       	e^{-i \vec k\cdot \vec x}
	    \right).
	\end{equation}
Its units are $\sqrt{(\text{energy})/(\text{length})}$.

\subsubsection{Interaction Hamiltonian}

We want to build a toy model for phononic interaction with external particles. The Hamiltonian we will consider is
\begin{equation}
\label{eq:interactionham}
	\op{H}_\mathrm{int}
=
	\frac{
		\charge}{2}
	\op{\phi}^2
	(\mathbf{\op{x}}),
\end{equation}
where $\charge$ is the ``charge'' of the interaction with units of length that we can, with foresight, use to define a dimensionless coupling constant
\begin{align}
    \alpha_\sonic \coloneqq \frac {\charge}  {\lambdabar_\particle},
\end{align}
in terms of a \textit{reduced sonic Compton wavelength} of the external particle\footnote{Note that we are free to define the numerical quantity $\lambdabar_\particle$ despite the fact that the external particle is not a field excitation.}
\begin{align}
    \lambdabar_\particle \coloneqq \frac{\hbar} {m c_\sonic}.
\end{align}
The cross section is evaluated by standard perturbation theory with the quantized interaction term
\begin{align}
\label{eq:interactinham2}
	\op{H}_\mathrm{int}
&
=
	\frac{\charge}{2}
	\frac{1}{L^{3}}
	\sum_{\vec k, \vec k'}
	    \frac{\hbar c_\sonic}{2 \sqrt{k k'}}
	\left(
		\op{a}_{\vec k}
		e^{i \vec k\cdot \op{\vec x}}
	+
		\op{a}^\dagger_{\vec k}
		e^{-i \vec k\cdot \op{\vec x}}
	\right)
	\nonumber \\*
& \quad \times
	\left(
		\op{a}_{\vec k'}
		e^{i \vec k' \cdot \op{\vec x}}
		+
		\op{a}^\dagger_{\vec k'}
		e^{-i \vec k' \cdot \op{\vec x}}
	\right)
	.
\end{align}
The Hamiltonian is discrete as we are considering quantization in a finite ``sonic universe'' of volume $L^3$.

Important to notice is that the interaction Hamiltonian in  Eq.~\eqref{eq:interactinham2} is not Lorentz invariant---neither with respect to light, nor sound---as the operator $\op{\vec x}$ cannot be applied in a relativistic scenario. Hence, it is valid only for external particles.
We will elaborate more in the next subsection, Sec.~\ref{Sec:InternalParticleCS}.

\subsection{Internal particle cross section}
\label{Sec:InternalParticleCS}

We are now equipped with the relevant mathematical machinery and physical understanding to compute scattering cross sections for phonon scattering experiments involving external particles. Before we proceed to do so, we will discuss the nature of the scattering cross section for phonon scattering experiments involving internal particles, as we will make qualitative comparisons between these two types of scattering experiments in what follows.

In Sec.~\ref{Sec:Phonon scattering from internal particles} we demonstrated that the kinematic description of phonon scattering from internal particles takes on the same mathematical form in both the laboratory frame and the comoving in-universe observer frame. In other words, we can refer to the kinematic description of phonon scattering from internal particles as being \textit{sonically Lorentz covariant}. The types of particles that we would expect to behave in this manner would be collective excitation quasiparticles~\cite{volovikUniverseHeliumDroplet2009,barceloElectromagnetismEmergentPhenomenon2014} belonging either to the analogue gravity medium itself (in addition to the phonons, which are also collective excitation quasiparticles) or to some other analogue gravity medium with the same characteristic speed of sound to which the phonon bearing medium is coupled.

Condensed-matter quantum field theory~\cite{lancasterQuantumFieldTheory2014} is the appropriate way to study collective excitation quasiparticles, and so it is a reasonable hypothesis that the appropriate quantum description of sonically relativistic particles would therefore be given by a sonically relativistic condensed matter field theory, analogous to quantum field theory but with the speed of sound taking the place of the speed of light. There is in fact evidence to support this hypothesis: for example, both Volovik~\cite{volovikUniverseHeliumDroplet2009} and Barcel\'{o} {\it et al}.~\cite{barceloElectromagnetismEmergentPhenomenon2014}\footnote{The latter work being based, in part, on the former.} have demonstrated the emergence of quantum electrodynamics as an effective dynamical description of certain collective excitation quasiparticles within condensed matter systems. We will therefore assume that this hypothesis is correct, and so---in analogy to quantum field theory---the dynamical equations of motion governing the scattering of phonons from internal particles will be sonically Lorentz covariant. In-universe observers' measurements of kinematics are \textit{also} sonically Lorentz covariant, and so the differential cross section for phonon scattering from internal particles will trivially be sonically Lorentz invariant from the perspective of in-universe observers, akin to how the differential cross section for actual Compton scattering (described by the \textit{Klein-Nishina formula}~\cite{peskinIntroductionQuantumField2018a,kleinNishinaFormula1929}) is actually Lorentz invariant.

When considering the differential cross section of phonon scattering from external particles we are not afforded the same convenience. We imagine our external particle to be some regular particle free of any inherent association to the analogue gravity medium: its classical description is merely that of a Newtonian particle, and its corresponding quantum description is given by regular quantum mechanics. We must therefore take a careful and considered approach in calculating the differential cross section of phonon scattering from external particles, and to this end we apply the same general procedure that we applied in calculating the comoving in-universe observer frame description of the kinematics.

\subsection{External particle cross section}

To compute the differential cross section of phonon scattering from external particles, we apply Eqs.~\eqref{Eq:CrossSectionDetailedDefinition} and~\eqref{Eq:TransitionRate} to the specific experimental scenario that we have in mind, which is discussed in Sec.~\ref{Sec:Paper_Strategy} and shown in Fig.~\ref{Fig:Compton Scattering not-boosted and boosted}.

\subsubsection{Initial and final states}

The quantum toy model we are considering is composed of two independent systems: the phonon and the external particle. Hence the initial and final states of the system can be written as follows:
\begin{align}
	\ket \init
=
	\ket {\vec p_\init}
\otimes
	\ket {\hbar \vec k_\init},
&&
	\ket \final
=
	\ket {\vec p_\final}
\otimes
	\ket {\hbar  \vec k_\final}
	.
\label{Eq:InitialAndFinalState}
\end{align}
$\ket {\vec p_{\smash{\init/\final}}}$ are the initial/final states of the external particle, and $\ket {\hbar  \vec k_{\smash{\init/\final}}}$ are the initial/final states of the phonon. In Table~\ref{Table:ParticlesInitialFinal} we illustrate the details of the states, where we have use the subscript $j = \init, \final$ for the final and initial states

\begin{table*}[t]
    \centering
\begin{tabular}{c c  c  c  c }
\toprule
 Particle Type
 & One-Particle State
 & Commutation Relations
 & Inner Product
 & One-Particle Identity
 \\
\midrule
 Phonon & $\phantom{\mathclap{\int_{\cube_L}d^3}} \ket {\hbar \vec k} = \op a^\dagger_{\vec k} \vacsonicket $
 & $\bigl[\op a_{\vec k },\op a^\dagger_{\vec k'}\bigr]= \delta_{\vec k,\vec k'}$
 & $\langle \hbar \vec k \ket {\hbar \vec k'} = \delta_{\vec k, \vec k' }$
 & $\mathbb{I} = \sum_{\vec k} \ket {\hbar \vec k} \langle \hbar \vec k | $\\
\addlinespace[0.7ex]
 External & $\ket {\vec p} =\frac{1}{L^{3/2}} \int_{\cube_L}d^3 x e^{\frac{i \vec p \cdot \vec x}{\hbar}} \ket {\vec x}$
 & N/A
 & $\langle  \vec p \ket {\vec p'} = \delta_{\vec p, \vec p' }$
 &  $\mathbb{I} = \sum_{\vec p} \ket {\vec p} \langle \vec p | $  \\
\bottomrule
\end{tabular}
    \caption{\label{Table:ParticlesInitialFinal}Representation of single-particle states of the phonon and external particle. The phonon is treated as single excitations (second quantized) of a sonically relativistic field---i.e., one with a dispersion relation $\omega = c_s k$. The external particle is treated as an ordinary quantum-mechanical particle (first quantized). Our choice of normalization allows all types of particle to have the same form of the inner product and resolution of the (one-particle) identity operator. Our choice of normalization for the phonon differs from that of the usual one in quantum field theory~\cite{peskinIntroductionQuantumField2018a}. This has no effect on observable quantities, including cross sections---see footnote~\ref{foot:norm}.
    }
\end{table*}
In line with the interaction Hamiltonian Eq.~\eqref{eq:interactinham2}, states are discrete as they are defined in a finite ``universe'' of volume $L^3$, and the delta functions are Kronecker deltas. Note that we are using a different normalization than what is usually used in literature, for example in Peskin~\cite{peskinIntroductionQuantumField2018a}. However, this does not have any effects on measured quantities (see footnote~\ref{foot:norm}.)

\subsubsection{Cross section derivation}

We now have all the general tools we need to calculate the differential and the total cross section in the laboratory frame and in the in-universe observers' reference frame. We will start with calculating the transition rate for Hamiltonian Eq.~\eqref{eq:interactinham2}, and the initial and final states in Table~\ref{Table:ParticlesInitialFinal}:
\begin{align}
	\bra \final \op H_\text{int}
	\ket \init
	.
\end{align}
If we plug in the expressions for the interaction Hamiltonian Eq.~\eqref{eq:interactinham2}, and initial and final states we obtain
\begin{align}
	\bra \final & \op H_\text{int} \ket \init
=
	\frac{\charge}{2}
	\bra {\vec p_\final}
\otimes
	\vacsonicbra
	\op a_{\vec k_\final}
	\nonumber
	\\*
&
\times
	\frac{1}{L^3}
	\sum_{\vec k, \vec k'}
	\frac{
	    \hbar
	    c_\sonic
	    }{
        2\sqrt{k k'}
        }
	\left(
		\op{a}_{\vec k}
		e^{i \vec k\cdot \op{\vec x}}
	+
		\op{a}^\dagger_{\vec k}
		e^{-i \vec k\cdot \op{\vec x}}
	\right)
	\nonumber
	\\*
&
\times
	\left(
		\op{a}_{\vec k'}
		e^{i \vec k' \cdot \op{\vec x} }
		+
		\op{a}^\dagger_{\vec k'}
		e^{-i \vec k' \cdot \op{\vec x}}
	\right)
	 \ket {\vec p_\init}
\otimes
	\op a^\dagger_{\vec k_\init}
	\vacsonicket
	.
\end{align}
By considering the commutation relations in Table~\ref{Table:ParticlesInitialFinal}, the orthogonality of the particle states, and well-known quantum mechanics relations,\footnote{For the full derivation see Appendix.} the transition amplitude can be rewritten as
\begin{align}
	\bra \final \op H \ket \init
=
	\frac{\charge}{2}
	\frac{\hbar c_\sonic}{L^3}
	\frac{1}
	    {
	    \sqrt{k_\init k_\final }}
			\delta_{\vec p_\final, \vec p_\init, + \hbar \vec k_\init - \hbar \vec k_\final}
\end{align}
Given the expression Eq.~\eqref{Eq:TransitionRate} we can rewrite the transition rate as
\begin{align}
	w_{\init\rightarrow \final}
&
=
	\frac{2 \pi}{\hbar}
	\abs{
		\frac{\charge}{2}
		\frac{\hbar c_\sonic}{L^3}
		\frac{1}{\sqrt{k_\init k_\final}}
		\delta_{\vec p_\final, \vec p_\init + \hbar \vec k_\init - \hbar \vec p_\final }
		}^2
		\delta(E_{\final \init})
		\nonumber
		\\*
&
=
	\frac{2 \pi}{\hbar}
	\frac{\charge^2}{4}
		\frac{(\hbar c_\sonic)^2}{L^6}
	\frac{1}{E_{k_\init} E_{k_\final}}
	\left(
		\delta_{\vec p_\final, \vec p_\init + \hbar \vec k_\init - \vec k_\final }
	\right)^2
	\delta(E_{\final \init})
	\nonumber
	\\*
&
=
	\frac{2 \pi}{\hbar}
	\frac{\charge^2}{4}
		\frac{(\hbar c_\sonic)^2}{L^6}
	\frac{1}{k_\init k_\final}
	\delta_{\vec p_\final, \vec p_\init + \hbar \vec k_\init - \vec k_\final }
	\delta(E_{\final \init})
	,
\label{Eq:TransitionRateExplicit}
\end{align}
where
\begin{align}
    E_{\final \init}
&
=
    E_\final
+
    \hbar\omega_\final
-
    E_\init
-
    \hbar\omega_\init
    \nonumber
    \\*
&
=
    E_\final (k_\final,k_\init, p_\init)
+
    \hbar c_\sonic k_\final
-
    E_\init
-
    \hbar c_\sonic k_\init
\end{align}
If we want to reexpress the cross section Eq.~\eqref{Eq:CrossSectionDetailedDefinition} in terms of continuous variables for the final state of the phonon, we can use the relation
\begin{align}
	\sum_{\vec k_\final}
&
=
	\frac{1}{(\hbar \Delta  k_\final)^3} \sum_{\vec k_\final} (\hbar \Delta  k_\final)^3
	\nonumber
	\\*
&
\approx
	\frac{1}{(\Delta k_\final)^3}
	\int_{\reals^3} d^3 k_\final
=
	\frac{L^3}{(2 \pi )^3}
	\int_{\reals^3} d^3 k_\final
	,
\end{align}
so that the total cross section is
\begin{align}
	\sigma[\mathcal{S}|\init]
=
	\frac{1}{\Phi}
	\frac{L^3}{(2 \pi)^3}
	\int_{\reals^3} d^3 k_\final
	\sum_{\vec p_\final}
	\frac{w_{\init\rightarrow \final}}{L^3}
	.
\label{Eq:CrossSectionContinuous}
\end{align}
Equation~\eqref{Eq:CrossSectionContinuous} is very useful to deduce the differential cross section with respect to the scattering angle $\cos \theta$
\begin{align}
	\frac{
		d \sigma
		}{
		d \cos \theta
		}
=
	\frac{1}{\Phi}
	\frac{L^3}{(2 \pi)^3}
	\int^{+ \infty}_{0} d k_\final \, k_\final^2
	\int^{2 \pi}_0 d \phi
	\sum_{\vec p_\final}
	\frac{w_{\init\rightarrow \final}}{L^3}
	.
\end{align}
Since we have assumed that the scattering processing we are considering is coplanar and colinear, we know that the quantity $w_{\init\rightarrow \final}$ does not depend on the angle $\phi$. Hence we can solve the integral in $\phi$ straightaway
\begin{align}
	\frac{
		d \sigma
		}{
		d \cos \theta
		}
=
	\frac{2 \pi}{\Phi}
	\frac{L^3}{(2 \pi)^3}
	\int^{+ \infty}_{0} d k_\final  \,  k_\final^2
	\sum_{\vec p_\final}
	\frac{w_{\init\rightarrow \final}}{L^3}
	.
\label{Eq:DifferentialCrossSectionGeneral}
\end{align}
Therefore the differential cross section Eq.~\eqref{Eq:DifferentialCrossSectionGeneral} becomes
\begin{align}
&
	\frac{
	    d \sigma[\mathcal{S}|\init]
	    }
	    {
	    d \cos \theta
	    }
=
	\frac{2 \pi }{\Phi}
	\frac{L^3}{(2 \pi)^3}
	\int^{+\infty}_{0} d k_\final k^2_\final
	\nonumber
	\\*
&
\times
		\sum_{\vec p_\final}
		\frac{1}{L^3}
		\frac{2 \pi}{\hbar}
		\frac{\charge^2}{4}
		\frac{(\hbar c_\sonic)^2}{L^6}
		\frac{1}{k_\init k_\final}
		\delta_{\vec p_\final, \vec p_\init + \hbar \vec k_\init - \hbar \vec k_\final}
		\delta(E_{\final \init})
	\nonumber
	\\*
&
=
	2 \pi
	\frac{L^6 }{\abs{c_\sonic - v_\init }}
	\frac{L^3}{(2 \pi)^3}
	\int^{+\infty}_{0} d k_\final k^2_\final
	\nonumber
	\\*
&
\times
		\sum_{\vec p_\final}
		\frac{1}{L^3}
		\frac{2 \pi}{\hbar}
		\frac{\charge^2}{4}
		\frac{(\hbar c_\sonic)^2}{L^6}
		\frac{1}{k_\init k_\final}
		\delta_{\vec p_\final, \vec p_\init + \hbar \vec k_\init - \hbar \vec k_\final}
		\delta(E_{\final \init})
	\nonumber
	\\*
&
=
	\frac{1}{\abs{1 - \beta_\init}}
	\frac{\hbar c_\sonic}{2\pi}
	\frac{\alpha^2_\sonic \lambdabar^2_\particle}{4}
	\nonumber
	\\*
&
\times
	\int^{+\infty}_{0} d k_\final k^2_\final
	\frac{1}{k_\init k_\final}
	\delta(E_{\final \init})
	\bigg|_{\vec p_\final = \vec p_\init + \hbar \vec k_\init - \hbar \vec k_\final}
	,
\label{Eq:CrossSectionGeneralDelta}
\end{align}
Equation~\eqref{Eq:CrossSectionGeneralDelta} can be further simplified by rewriting the Dirac delta as
\begin{align}
	\delta(E_{\final \init})
=
	\sum_{ k_0 \in \ker(E_{\final \init})}
		\frac{
			\delta(
				 k_\final
			-
				 k_0
				)
			}{
			\abs{
				(d E_{\final \init} / d k_\final)_{k_\final = k_0}
				}
			}
	,
\end{align}
where $\ker(E_{\final \init}) \coloneqq \{ k_0:  E_{\final \init} (k_0)= 0\}$, and as before, $E_{\final\init}$ is the analytic expression for the change in the total energy of the system (i.e.,~the difference between the final and the initial energies of the system).
The differential cross section Eq.~\eqref{Eq:DifferentialCrossSectionGeneral} is
\begin{align}
	\frac{d \sigma}{d \cos \theta}
&
=
	\frac{1}{\abs{1 - \beta_\init}}
	\frac{\hbar c_\sonic}{(2\pi)}
	\frac{\alpha^2_\sonic \lambdabar^2_\particle}{4}
	\int^\infty_{0} d k_\final \, k^2_\final
	\nonumber
	\\*
&
\times
	\frac{1}{k_\init k_\final}
	\sum_{ k_0 \in \ker(E_{\final \init})}
	\frac{
		\delta(
			 k_\final
		-
			 k_0
			)
		}{
		\abs{
			(d E_{\final \init} / d k_\final)_{k_\final = k_0}
			}
		}
	\bigg|_{\vec p_\final = \vec p_\init + \hbar \vec k_\init - \hbar \vec k_\final}
	\label{Eq:DifferentialCrossSectionGeneralDelta}
\end{align}

\subsubsection{Scattering cross sections in the laboratory frame}

We can now proceed to derive the differential cross section for phonon scattering from external particles in the laboratory frame. From the kinematic derivation in Sec.~\ref{SubSec:ExternaParticleScatteringLab} we know that we have two solutions $\hbar k_{\final,1}$ and $\hbar k_{\final, 2}$, see Eq.~\eqref{Eq:FinalPhononEnergy}, that satisfy the condition $E_{\final \init} = E^\text{tot}_\final - E^\text{tot}_\init = 0$, see Eq.~\eqref{Eq:ConserEnergyExternal1}. The differential cross section Eq.~\eqref{Eq:DifferentialCrossSectionGeneralDelta} becomes
\begin{align}
	\frac{d \sigma[\mathcal{S}|\init]}{d \cos \theta}
&
=
	\frac{1}{\abs{1 - \beta_\init}}
	\frac{\hbar c_\sonic}{(2\pi)}
	\frac{\alpha^2_\sonic \lambdabar^2_\particle}{4}
	\frac{1}{k_\init}
	\nonumber
	\\*
&
\times
	\int^{+ \infty}_{0}
		d k_\final \,  k_\final
		\sum_{k_0 \in \left\{k_{\final,1}, k_{\final,2}\right\}}
			\frac{
				\delta(
					k_\final
				-
					k_0
					)
				}{
				\abs{
					(d E_{\final \init} / d k_\final)_{k_\final = k_0}
					}
				}
	.
\end{align}
The denominator is easily calculated from Eq.~\eqref{Eq:ConserEnergyExternal1}
\begin{align}
	\frac{d E_{\final \init}}
	{d k_\final}
=
	\frac{\hbar }{m}
	\left[
		\hbar k_\final
	-
		\left(
			\hbar k_\init
		+
			p_\init
		\right)
		\cos \theta
	+
		m c_\sonic
	\right]
	.
\end{align}
By expanding the sum and considering the two solutions of $\ker(E_{\final \init})$ the differential cross section becomes
\begin{align}
&
	\frac{d\sigma[\mathcal{S}|\init]}{d \cos \theta}
=
	\frac{1}{\abs{1 - \beta_\init}}
	\frac{\hbar c_\sonic}{(2\pi)}
	\frac{\alpha^2_\sonic \lambdabar^2_\particle}{4}
	\frac{1}{k_\init}
	\int^{+ \infty}_{0}
	d k_\final \, k_\final
	\nonumber
	\\*
&
\times
	\left[
		\frac{\delta(k_\final-k_{\final,1})
		    }{
			\abs{
				\frac{1}{m}
				\hbar
				\left[
				    \hbar
					k_{\final,1}
				-
					\left(
						\hbar
						k_\init
					+
						p_\init
					\right)
					\cos \theta
				+
					m c_\sonic
				\right]
			}
		}
	\right.
	\nonumber
	\\*
&
+
    \left.
		\frac{
		    \delta(k_\final- k_{\final,2})
		    }{
			\abs{
				\frac{1}{m}
				\hbar
				\left[
					\hbar k_{\final,2}
				-
					\left(
						\hbar k_\init
					+
						p_\init
					\right)
					\cos \theta
				+
				m c_\sonic
			    \right]
			    }
		    }
	\right]
\end{align}
When we perform the integral in $k_\final$ we find
\begin{align}
&
	\frac{d \sigma[\mathcal{S}|\init]}{d \cos \theta}
=
	\frac{1}{\abs{1 - \beta_\init}}
	\frac{\hbar c_\sonic}{(2\pi)}
	\frac{\alpha^2_\sonic \lambdabar^2_\particle}{4}
	\frac{1}{k_\init}
	\frac{m}{\hbar}
	\nonumber
	\\*
&
\times
	\frac{
		(k_{\final,1} H[k_{\final,1}] + k_{\final,2} H[k_{\final,2}])
		}{
		\sqrt{
			\bar B^2
		-
			\bar C
		}
	}
	,
\label{Eq:DifferentialCrossSectionExternalFinal}
\end{align}
where $\bar B$ and $\bar C$ are defined in Eq.~\eqref{eqs:BCbar}. The functions $ H[k_{\final,1}]$ and $H[k_{\final,2}]$ are the Heaviside step functions that ensure that the two solutions $k_{\final,1}$ and $k_{\final,2}$ are real and positive, as we have already seen in the kinematic Sec.~\ref{SubSec:ExternaParticleScatteringLab}. By inserting the two solutions from Eq.~\eqref{Eq:FinalPhononEnergy} into Eq.~\eqref{Eq:DifferentialCrossSectionExternalFinal}, the explicit expression for the differential cross section is
\begin{align}
&
	\frac{d \sigma[\mathcal{S}|\init]}{d \cos \theta}
=
	\frac{1}{\abs{1 - \beta_\init}}
	\frac{\hbar c_\sonic}{(2\pi)}
	\frac{\charge^2}{4}
	\frac{1}{k_\init}
	\frac{m}{\hbar^2}
	\nonumber
	\\*
&
\times
	\left[
	    2
	    A
	    H[\bar B -\sqrt{\bar C}]
		H[\bar C]
	+
	    \left(
	        A
	   +
	        1
	    \right)
	     H(-\bar C)
	\right]
	,
\label{Eq:DifferentialCrossSectionExternalFinal2}
\end{align}
where we have defined
\begin{align}
    A
\coloneqq
    \frac{\bar B}
        {
        \sqrt{
 			\bar B^2
        -
 			\bar C
 		}}
 	.
\end{align}
The three conditions specified by the three Heaviside step functions $H(\bar B-\sqrt{\bar C})$,  $H(\bar C)$, and $H(-\bar C)$ are the same as those in Eqs.~\eqref{Eq:Condition1}. %
The differential cross section has units of area, as it must, since the quantity $\frac{\hbar c_\sonic}{k_\init} \frac{m}{\hbar^2}$ is dimensionless.

\subsubsection{Scattering cross sections in the comoving in-universe observer frame}

As we have seen multiple times throughout this paper, the comoving in-universe observer frame description of the scattering event is obtained by appropriately transforming the laboratory frame kinematic quantities into the comoving frame of the particle. %
We have already performed all of the heavy lifting required to obtain the differential cross section in the frame of the comoving in-universe observer.
Our final task is to transform the kinematic quantities in Eq.~\eqref{Eq:DifferentialCrossSectionExternalFinal} [or, alternatively, Eq.~\eqref{Eq:DifferentialCrossSectionExternalFinal2}] according to the transformations specified in Eqs.~\eqref{Eq:Transformation for initial frequency}--\eqref{Eq:TransformationCosTheta}. Substituting $k_\init = D k'_\init$,
where $D$ is the Doppler factor [Eq.~\eqref{Eq:Doppler_factor}], and applying Eq.~\eqref{Eq:DiffCSGeneralTransformation},
the differential cross section becomes
\begin{align}
&
	\frac{
		d \sigma[\mathcal{S}'|\init']
		}{
		d \cos \theta'
		}
=
	\frac{1+\beta_\init}{(1+\beta_\init \cos \theta')^2}
	\frac{\charge^2 \hbar c_\sonic}{4(2\pi)}
	\frac{m}{\hbar^2}
	\frac{1}{D k'_\init}
	\nonumber
	\\*
&
\times
	\left[
	    2
        A'
        H(\bar B'-\sqrt{\bar C'})
		H(\bar C')
	+
	    \left(
	        A'
	   +
	        1
	    \right)
	    H(-\bar C')
	\right]
	,
\label{Eq:DiffCrossSectionInUniverseExternal}
\end{align}
where $\bar B'$ and $\bar C'$ are just $\bar B$ and $\bar C$ reexpressed in the comoving frame, as shown in Eq.~\eqref{eqs:BCbarprime}, and $A'$ is
\begin{equation}
    A'
=
    \frac{\bar B'}{
    \sqrt{\bar B'^2 - \bar C'}}
    .
\end{equation} %
\subsubsection{Using Lorentz-violating sonic Compton scattering to determine absolute motion}
\label{Subsec:Using Lorentz-violating sonic Compton scattering to determine absolute motion}

The differential cross section written in terms of the values of quantities as measured by comoving in-universe observers is given by Eq.~\eqref{Eq:DiffCrossSectionInUniverseExternal}. In principle, in-universe observers could use this expression to make qualitative statements regarding their state of motion by performing several scattering experiments on external particles with different initial velocities $v_\init  = \beta_\init c_\sonic$, using phonons with different initial energies $\hbar \omega'_\init$ (or equivalently $\hbar c_\sonic k'_\init$). To facilitate the comparison between different energy and velocity regimes, we introduce the dimensionless quantity
\begin{equation}
    \zeta'
=
    \frac{
        \hbar \omega'_\init
        }{
        mc^2_\sonic}
=
    \frac{
        \hbar k'_\init
        }{
        m c_\sonic}
=
    \lambdabar_\particle
    k'_\init
    ,
\end{equation}
that represents the ratio of the initial energy of the phonon to the ``sonic rest-mass energy'' that in-universe observers would think to associate to the particle.

The ratio $\zeta^\prime$ is directly analogous to a ratio that appears in the description of actual Compton scattering, which we shall here explicitly denote\footnote{The expression for $\xi$ is identical in form to that of $\zeta^\prime$ except all references to sound are replaced by their corresponding references to light. That is to say, $c$ takes the place of $c_\sonic$, and the frequencies $\omega^\prime_\init$ and $\omega^\prime_\final$ correspond to frequencies of light, not sound.} $\xi$, the value of which has particular physical implications. The limit $\xi\ll{1}$ corresponds to Thomson scattering~\cite{itzyksonQuantumFieldTheory2012}, which describes the classical and nonrelativistic scattering of electromagnetic waves from charged particles, whereas the limit $\xi\gtrsim{1}$ corresponds to scenarios in which both relativistic and quantum theoretic effects are prominent: as a result, a full quantum field theoretic description is necessary in the limit $\xi\gtrsim{1}$~\cite{diracRelativisticQuantumMechanics1932}.

In the following we demonstrate what in-universe observers will measure if they were to perform several scattering experiments for the same value of $\zeta'$ and several values of $\beta_\init$ and vice versa. We provide the differential cross sections as viewed from the comoving frame for two types of experiments: in one type of experiment, the value of $\zeta'$ is held constant while $\beta_\init$ is varied; in the other type of experiment, $\beta_\init$ is held constant while $\zeta'$ is varied.

\begin{figure}[tb]
    \centering
        \includegraphics[width=1\columnwidth]{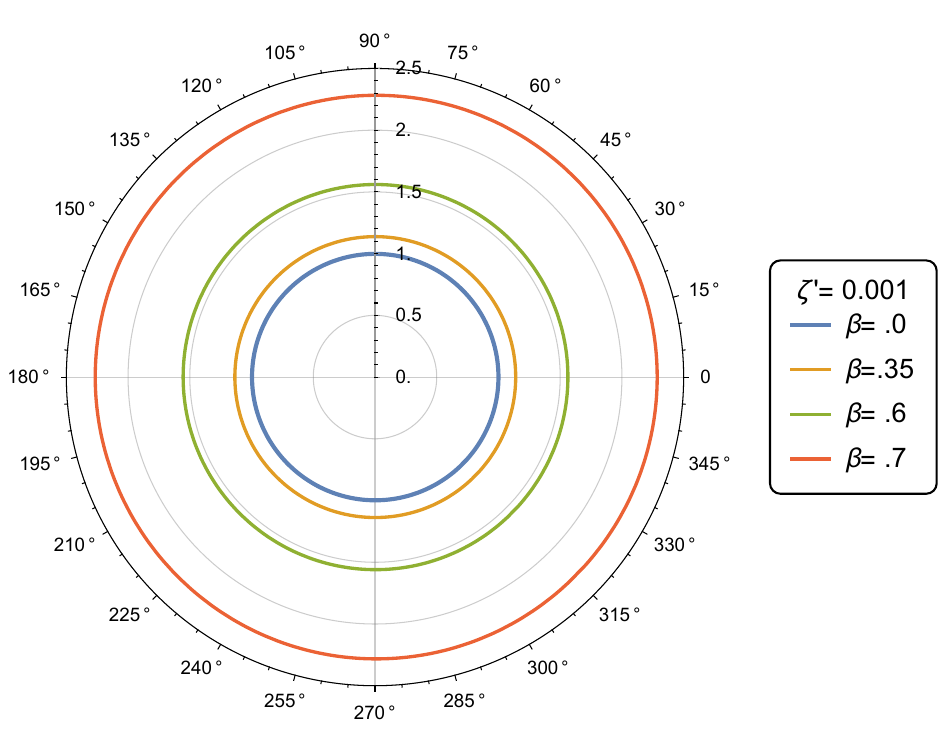}
        \caption{Polar plot of the differential cross section  for an external particle in the in-universe comoving frame for $\zeta' = 0.001$ and various values of $\beta$.}
        \label{fig:ComDiffCSbetaszeta0001}
\end{figure}

\begin{figure}[tb]
    \centering
        \includegraphics[width=1\columnwidth]{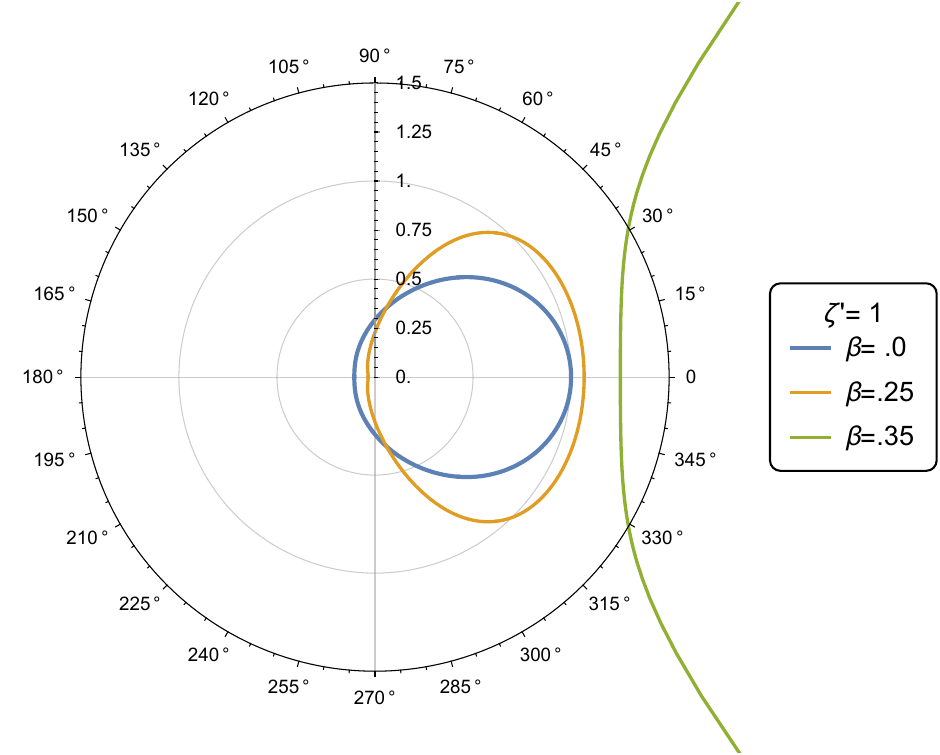}
        \caption{Polar plot of the differential cross section for an external particle in the in-universe comoving frame for $\zeta' = 1$ and various values of $\beta$.}
        \label{fig:ComDiffCSbetaszeta1}
\end{figure}

\begin{figure}[tb]
    \centering
        \includegraphics[width=1\columnwidth]{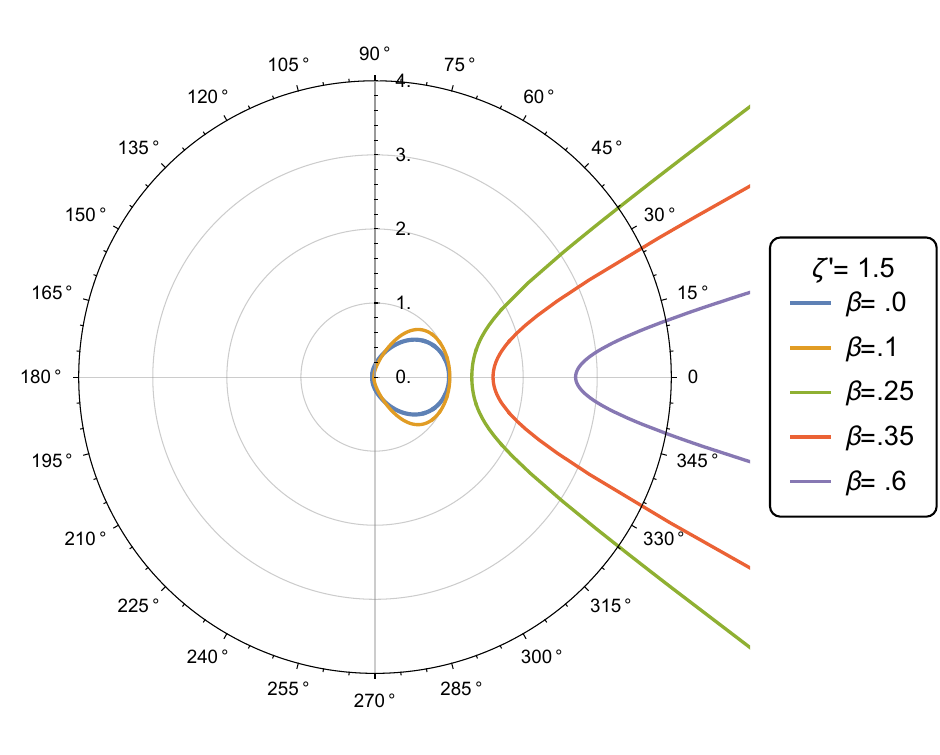}
        \caption{Polar plot of the differential cross section for an external particle in the in-universe comoving frame for $\zeta' = 1.5$ and various values of $\beta$.}
        \label{fig:ComDiffCSbetaszeta1p5}
\end{figure}

\paragraph*{Fixed $\zeta'$, varying $\beta_\init$.---} \hspace{-4mm} For this case we present the scattering cross sections for $\zeta'=\lbrace{0.001,1,1.5}\rbrace$. In Fig.~\ref{fig:ComDiffCSbetaszeta0001} we can already see that, even for a very small value of $\zeta'$, the cross section shows a marked dependence on different values of $\beta_\init$. We can qualitatively compare this result with the Klein-Nishima differential cross section formula~\cite{kleinNishinaFormula1929,weinbergElectrodynamics1995} for unpolarized photons, where it is easy---and expected---to see that there is no dependency of the cross section on the state of motion of the rest frame of the particle (for in actual relativity, there is no meaningful notion in which different inertial states of motion differ). For higher values of $\zeta'$, another noteworthy feature becomes apparent: in Figs.~\ref{fig:ComDiffCSbetaszeta1} and~\ref{fig:ComDiffCSbetaszeta1p5} we can see that, as the initial velocity of the particle increases, the range of the scattering angle becomes smaller, eventually leading to a scenario in which the particle is prohibited from scattering outside of some given angular window. This effect is due to the conditions in Eq.~\eqref{Eq:RealPositiveConditionsInUniverse}, which place restrictions on the allowed angles of scattering as per Eq.~\eqref{Eq:Comoving_angular_restrictions}.


\begin{figure}[tb]
    \centering
        \includegraphics[width=1\columnwidth]{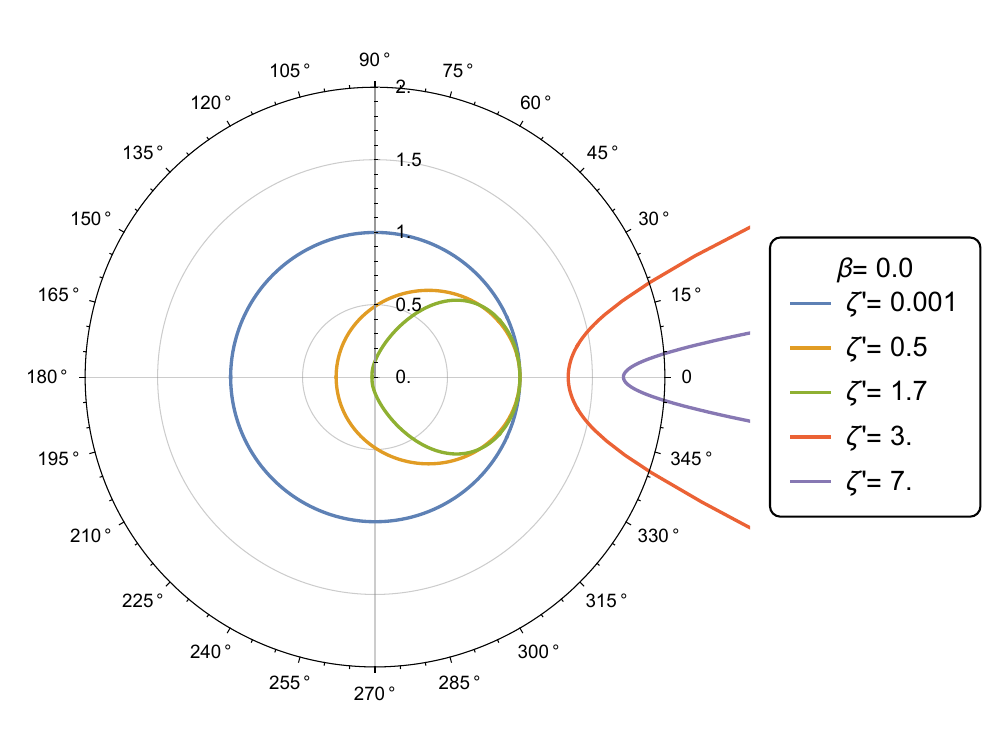}
        \caption{Polar plot of the differential cross section for an external particle in the in-universe comoving frame for $\beta_\init = 0$ and various values of $\zeta$.}
        \label{fig:ComDiffCSzetasbeta0}
\end{figure}

\begin{figure}[tb]
    \centering
        \includegraphics[width=1\columnwidth]{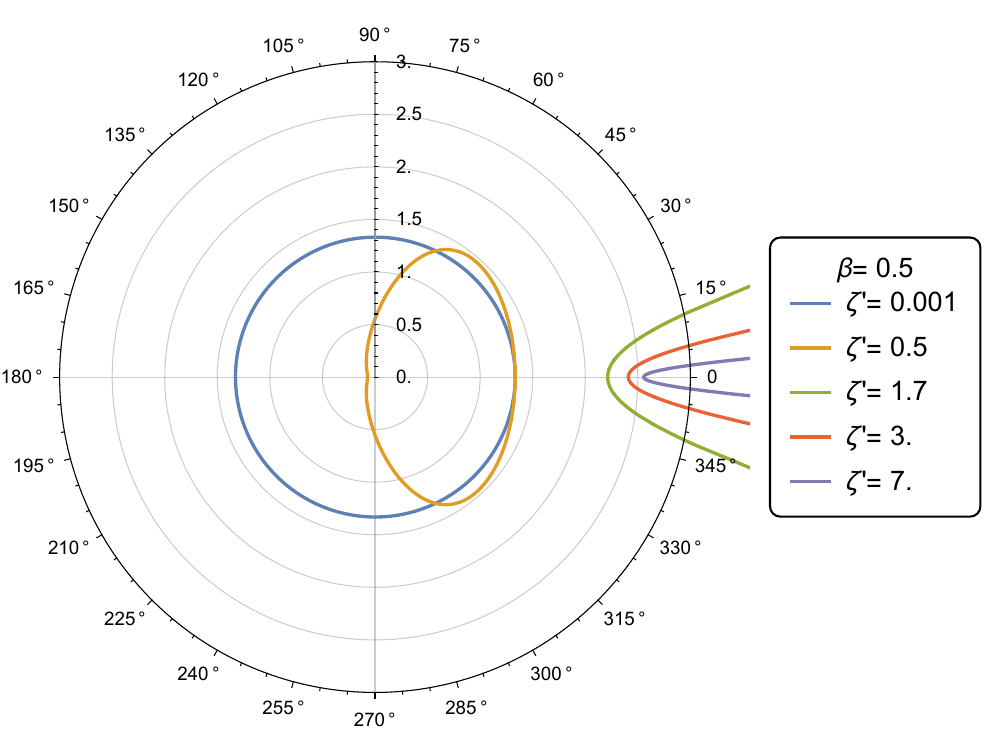}
        \caption{Polar plot of the differential cross section for an external particle in the in-universe comoving frame for $\beta = 0.5$ and various values of $\zeta'$.}
        \label{fig:ComDiffCSzetasbeta05}
\end{figure}

\paragraph*{Fixed $\beta_\init$, varying $\zeta^\prime$.---} \hspace{-3mm}The dependency of the scattering angle on the initial energy of the phonon $\zeta'$ becomes clearer when we consider the in-universe differential cross section for two values of $\beta_\init = \lbrace{0, 0.5\rbrace}$ and several values of $\zeta'$. In  Figs.~\ref{fig:ComDiffCSzetasbeta0} and~\ref{fig:ComDiffCSzetasbeta05} we can see that as $\zeta'$ increases, the scattering angle becomes more forward, tending towards the direction of motion of the particle in the laboratory frame. This effect becomes more pronounced with increasing $\beta_\init$, as one can see by comparing Figs.~\ref{fig:ComDiffCSzetasbeta0} and \ref{fig:ComDiffCSzetasbeta05}.


\section{Discussion}
\label{Sec:Discussion}

In considering phonon scattering from \textit{external particles} we have restricted our considerations to that of nonrelativistic quantum mechanics for all values of $\zeta^\prime$ because, by construction, our external particle is a nonrelativistic quantum mechanical object. By analogy to true Compton scattering, phonon scattering from \textit{internal particles} requires a full quantum field theoretic description\footnote{As previously discussed, a condensed matter quantum field theory.} because relativistic effects (with respect to sound) are important, and relativistic quantum mechanics is inapplicable because it is a fundamentally inconsistent theory~\cite{diracRelativisticQuantumMechanics1932}. We can however make qualitative comparisons between phonon scattering from internal particles and external particles in the limit $\zeta^\prime\ll{1}$, as in this limit relativistic effects become unimportant for internal particles. Strictly speaking, any comparisons of this type should be made in the specific case for which both types of particle are initially traveling very slowly in the laboratory frame ($\beta_\init\ll1$) as it is in this limit that the energy-momentum relations for internal and external particles coincide and thus the limit in which a quantum field theoretic description should coincide with an ordinary quantum mechanical description. With this in mind, Fig.~\ref{fig:ComDiffCSbetaszeta0001} shows the differential cross sections for phonon scattering from external particles characterized by $\zeta^\prime=10^{-3}$: the limit $\zeta^\prime\ll{1}$ is respected here, and while the total amount of scattering is a function of $\beta_\init$, the overall form of the differential scattering cross section (i.e.,~its angular dependency) is not. That the angular dependency of the differential scattering cross section is insensitive to $\beta_\init$ in this case is what we expect from the expected equivalence between phonon scattering from internal and external particles for low $\zeta^\prime$.

Any qualitative similarities between phonon scattering from internal and external particles vanishes for higher values of $\zeta^\prime$. In Figs.~\ref{fig:ComDiffCSbetaszeta1} and~\ref{fig:ComDiffCSbetaszeta1p5} the differential scattering cross sections for various values of $\beta_\init$ are plotted for $\zeta^\prime=1$ and $\zeta^\prime=1.5$, respectively. In both cases not only does the amount of scattering vary with initial $\beta_\init$, but the overall form of the differential scattering cross sections is sensitive to changes in $\beta_\init$ too: this is in stark contrast to what would occur for scattering from internal particles which, again, must be insensitive to the value of $\beta_\init$ due to sonic Lorentz covariance that is inbuilt into internal particles.

The effect of varying $\zeta^\prime$ for constant values of $\beta_\init$ \textit{should} be expected to alter the angular distribution of scattering, even for internal particles. In real Compton scattering, forward scattering becomes preferentially favored as the initial photon energy is increased (see, for example,~\cite{nelmsGraphsComptonEnergyangle1953}). The fact that there exists angular dependency in the differential scattering cross section of phonons from external particles for fixed $\beta_\init$ in Figs.~\ref{fig:ComDiffCSzetasbeta0} and~\ref{fig:ComDiffCSzetasbeta05} is therefore not in and of itself surprising or unexpected. With that said, for phonon scattering from external particles, increasing $\zeta^\prime$ for fixed $\beta_\init$ eventually reveals the presence of forbidden scattering angles: this has \textit{no} qualitative similarity to Compton scattering, and thus no qualitative similarity to phonon scattering from internal particles.

\section{Conclusion}
\label{Sec:Conclusion}

Provided that in-universe observers in an analogue-gravity universe are allowed to interact with Newtonian particles external to their own medium, then the sonic analogue to Compton scattering---in which phonons scatter from these external particles---can be used by in-universe observers to infer that there must exist some preferred rest frame.	In all but the most restrictive cases (i.e.,~unless $\beta\approx0$ and $\zeta^\prime\ll1$), external particles result in qualitatively different scattering profiles than occur in fully relativistic scattering, such as Compton scattering. In the most dramatic cases, scattering from external particles results in differential scattering cross sections with forbidden angles, and as the energy of the phonon increases relative to the ``sonic rest-mass energy'' of the particle ($\zeta^\prime$ increases) the window of allowed scattering angles becomes more tightly concentrated in the direction of the trajectory of the particle prior to scattering. Even in the cases for which scattering occurs at all angles, scattering for fixed values of $\zeta^\prime$ shows a preference towards forward scattering for increasing $\beta_\init$.

In principle, in-universe observers could conceivably utilize phonon scattering experiments from external particles to identify not only that a preferred rest frame must exist, but specifically which frame is the rest frame of their analogue universe. The ability to resolve which frame is actually the laboratory frame is fundamentally constrained by the mass of external particles and the energies of phonons that in-universe observers have access to. If in-universe observers are only able to probe the parts of parameter space corresponding to $\zeta^\prime\ll1$ then the angular distribution of scattering will not be considerably affected by their state of velocity (see Fig.~\ref{fig:ComDiffCSbetaszeta0001}) and thus they will only be able to detect their state of motion provided that they correctly deduce the relationship between $\beta_\init$ and the magnitude of the differential scattering cross section (that is, lower magnitudes correspond to lower $\beta_\init$, as per Fig.~\ref{fig:ComDiffCSbetaszeta0001}). If, on the contrary, in-universe observers are able to probe regions of parameter space corresponding to $\zeta^\prime\gtrsim{1}$, then the presence of forbidden scattering angles could be utilized to locate the medium's rest frame. In order for in-universe observers to utilize forbidden scattering angles to their advantage, they would either have to reverse engineer the energy-momentum relation for external particles, or postulate the correct energy-momentum relation and confirm it experimentally. Given that the correct energy-momentum relation for external particles corresponds to the $\beta_\init\ll1$ limit of internal particles, it is not unreasonable to think that they would eventually postulate the correct relation.

	 Scattering experiments performed from internal particles must ultimately be equivalent to scattering experiments performed within a truly relativistic theory due to the fact that a Lorentz symmetry (with respect to $c_\sonic$) is inbuilt into both internal particles themselves and the reference frames of in-universe observers. The interpretation here is that when internal particles are used, every part of the system is sonically Lorentz covariant, and when this is true the whole analogue-gravity system can be treated as being a sonic analogue to something like Lorentz ether theory, which is operationally indistinguishable from special relativity. Thus, when in-universe observers are only allowed to interact with internal particles, they cannot determine the presence of a preferred rest frame. It is only when the symmetry groups obeyed by the particle and the phonons are different that the analogy between our model and a Lorentz ether theory (and hence a relativistic theory) breaks down and the presence of the medium can be detected.

	As a final note, to in-universe observers the sonic analogue to Compton scattering from external particles constitutes a breaking of the sonic Lorentz symmetry that they would otherwise believe in without access to external objects. From this point of view, the Standard Model extension~\cite{colladayLorentzviolatingExtensionStandard1998} might provide a natural way to further investigate such scenarios from the perspective of in-universe observers who want to believe that the sonic Lorentz symmetry of their universe is fundamental. To highlight this point, consider that an excitation of a truly relativistic field (that is, relativistic with respect to the speed of light) would also constitute an example of an external particle from the perspective of in-universe observers. The Lagrangian density $\mathcal{L}_l$ describing a real scalar field that is actually relativistic (the subscript $l$ denotes that the Lagrangian is invariant under Lorentz transformations with respect to $c$ the speed of light) can be written in terms of some sonically relativistic Lagrangian $\mathcal{L}_s$ (the subscript $s$ denotes that the Lagrangian is invariant under Lorentz transformations with respect to $c_\sonic$ the speed of sound) with some additional term $\mathcal{K}$ to account for the difference:
	\begin{equation}
	    \mathcal{L}_l = \mathcal{L}_s + \mathcal{K}.
	\end{equation}
	Viewed this way, in-universe observers might be able to describe certain external particles as though they were external particles that were coupled to some background vector field defining a preferred frame (the rest frame of the medium). This is precisely the type of scenario that the standard model extension deals with, though in the standard model extension the Lorentz symmetry is taken to be fundamental at high energies and spontaneously broken at low energies, whereas the sonic Lorentz symmetry is emergent rather than fundamental.

    \acknowledgments
	We thank Sundance Bilson-Thompson for useful discussions and for the original suggestion to investigate the sonic analogue to Compton scattering as a starting point for investigating more general phenomenon in analogue-gravity settings. We also thank Carlos Barcel\'{o} for his insightful discussions in the early stages of this project, and for the hospitality that he extended to S.L.T.\ and G.P.\ during a visit to the beautiful city of Granada, Spain. This work was supported by the U.S. Air Force Research Laboratory (AFRL) Asian Office of Aerospace Research and Development (AOARD) under Grant Number FA2386-16-1-4020, by the Australian Research Council~(ARC) Discovery Program (Project No.\ DP200102152), and by the ARC Centre of Excellence for Quantum Computation and Communication Technology (Project No.\ CE170100012).

	\newpage
	\appendix

	\begin{widetext}
	\section{General transition probability derivation}
	\label{App:CrossSectionDerivation}

	From the definition of cross section in Eq.~\eqref{Eq:CrossSectionDetailedDefinition}, or equivalently from Eq.~\eqref{Eq:CrossSectionContinuous}, we see that we need to evaluate the amplitude
	\begin{equation}
	\label {Eq:PAD}
		\bra \final {\op H_\text{int} } \ket \init,
	\end{equation}
	where the initial and final states are, respectively,
	\begin{align}
		\ket \init
	=
		\ket{\vec p_\init}
	\otimes
		\ket{\hbar  \vec k_\init}
		,
	\qquad
		\ket \final
	=
		\ket{\vec p_\final}
	\otimes
		\ket{\hbar  \vec k_\final}
		,
	\end{align}
	and where the states for external particles are defined in Table~\ref{Table:ParticlesInitialFinal}. We consider the states represented in momentum space, so the phonon initial and final states can be written as
	\begin{align}
	    \ket{\hbar \vec k_\init}
	=
	    \op a^\dagger_{\vec k_\init}\vacsonicket
	    ,
	    \qquad
	    \ket{\hbar \vec k_\final}
	=
	    \op a^\dagger_{\vec k_\final}\vacsonicket
	    ,
	\end{align}
	where $\vacsonicket$ is the phonon ground state.
	The amplitude Eq.~\eqref{Eq:PAD} then becomes
	\begin{align}
		\bra \final {\op H_\text{int} } \ket \init
	&
	=
		\frac{\charge}{2}
		\frac{1}{L^3}
		\sum_{\vec k, \vec k'}
		\frac{\hbar c_\sonic}{2\sqrt{k k'}}
	    \bra{\vec p_\final}
	\otimes
		\vacsonicbra
		\op a_{\vec k_\final}
		\left(
			\op{a}_{\vec k}
			e^{i \vec k\cdot \op{\vec x}}
		+
			\op{a}^\dagger_{\vec k}
			e^{-i \vec k\cdot \op{\vec x}}
		\right)
		\left(
			\op{a}_{\vec k'}
			e^{i \vec k' \cdot \op{\vec x} }
			+
			\op{a}^\dagger_{\vec k'}
			e^{-i \vec k' \cdot \op{\vec x}}
		\right)
		 \ket{\vec p_\init}
	\otimes
		\op a^\dagger_{\vec k_\init}
			\vacsonicket
			\nonumber
			\\*
	&
	=
		\frac{\charge}{2}
		\frac{1}{L^3}
		\sum_{\vec k, \vec k'}
		\frac{\hbar c_\sonic}{2\sqrt{k k'}}
		\bra{\vec p_\final}
	\otimes
		\vacsonicbra
		\op a_{\vec k_\final}
		\left(
			\op{a}_{\vec k}
			\op{a}^\dagger_{\vec k'}
			e^{i
			(\vec k - \vec k')
			\cdot
			 \op{\vec x} }
		+
			\op{a}^\dagger_{\vec k}
			\op{a}_{\vec k'}
			e^{-i (\vec k - \vec k')
			\cdot
			\op{\vec x}}
		\right)
		 \ket{\vec p_\init}
	\otimes
		\op a^\dagger_{\vec k_\init}
		\vacsonicket
		\nonumber
		\\*
		&
	=
		\frac{\charge}{2}
		\frac{\hbar c_\sonic}{L^{3}}
		\sum_{\vec k, \vec k'}
		\frac{1}{2\sqrt{k k'}}
		\bra{\vec p_\final}
	\otimes
		\vacsonicbra
		\op a_{\vec k_\final}
		\Bigl(
			\op{a}_{\vec k}
			\op{a}^\dagger_{\vec k'}
			e^{i
			(\vec k - \vec k')
			\cdot
			 \op{\vec x} }
		+
			\underbrace{
			\op{a}^\dagger_{\vec k'}
			\op{a}_{\vec k}
			e^{-i (\vec k' - \vec k)
			\cdot
			\op{\vec x}}
			}_{(*)}
		\Bigr)
		 \ket{\vec p_\init}
	\otimes
		\op a^\dagger_{\vec k_\init}
		\vacsonicket
		.
	\end{align}
	In the term $(*)$ we have swapped $\vec k$ and $\vec k'$ as these are dummies variables. Using the commutation relations mentioned in Table~\ref{Table:ParticlesInitialFinal} we obtain
	\begin{align}
		\bra \final {\op H_\text{int} } \ket \init
		&
	=
		\frac{\charge}{2}
		\frac{\hbar c_\sonic}{L^3}
		\sum_{\vec k, \vec k'}
		\frac{1}{2\sqrt{k k'}}
		\bra{\vec p_\final}
	\otimes
		\vacsonicbra
		\op a_{\vec k_\final}
		\left(
			2 \op{a}^\dagger_{\vec k'}
			 \op{a}_{\vec k}
		+
		\delta_{\vec k, \vec k'}
		\right)
		e^{i (\vec k - \vec k')
		\cdot
		\op{\vec x}}
		\ket{\vec p_\init}
	\otimes
		\op a^\dagger_{\vec k_\init}
		\vacsonicket
		\nonumber
		\\*
	&
	=
		\frac{\charge}{2}
		\frac{\hbar c_\sonic}{L^3}
		\sum_{\vec k, \vec k'}
		\frac{1}{2\sqrt{k k'}}
		\Biggl(
			\underbrace{
			\bra{\vec p_\final}
		\otimes
			\vacsonicbra
			\op a_{\vec k_\final}
				2 \op{a}^\dagger_{\vec k'}
			 	\op{a}_{\vec k}
			e^{i (\vec k - \vec k')
			\cdot
			\op{\vec x}}
			\ket{\vec p_\init}
		\otimes
			\op a^\dagger_{\vec k_\init}
			\vacsonicket}_{(**)}
		+
			\underbrace{
			\bra{\vec p_\final}
		\otimes
			\vacsonicbra
			\op a_{\vec k_\final}
			\delta_{\vec k, \vec k'}
			e^{i (\vec k - \vec k')
			\cdot
			\op{\vec x}}
			\ket{\vec p_\init}
		\otimes
			\op a^\dagger_{\vec k_\init}
			\vacsonicket}_{(***)}
		\Biggr)
		.
	\end{align}
	We evaluate $(**)$ first:
	\begin{align}
		(**)
	=
		2 \, \bra{\vec p_\final}
		e^{i (\vec k - \vec k')
		\cdot
		\op{\vec x}}
		\ket{\vec p_\init}~
		\vacsonicbra
		\op a_{\vec k_\final}
		\op{a}^\dagger_{\vec k'}
		\op{a}_{\vec k}
		\op a^\dagger_{\vec k_\init}
		\vacsonicket
		.
	\end{align}
	Considering that $\op a_{\vec k}\op{a}^\dagger_{\vec k'} =\op{a}^\dagger_{\vec k'}   \op a_{\vec k} + \delta_{\vec k,\vec k'}$, and that $e^{i \vec k \cdot  \op{\vec x}} \ket{\vec p} = \ket{\vec p + \hbar \vec k}$, $(**)$ becomes
	\begin{align}
		(**)
	&
	=
		2  \inprod{\vec p_\final}
		{\vec p_\init + \hbar \vec k - \hbar \vec k'} \, \,
		\vacsonicbra
		\left(
			\op{a}^\dagger_{\vec k'}
			\op a_{\vec k_\final}
			+
			\delta_{\vec k_\final ,\vec k'}
		\right)
		\left(
			\op a^\dagger_{\vec k_\init}
			\op{a}_{\vec k}
			+
			\delta_{\vec k,\vec k_\init }
		\right)
		\vacsonicket
		\nonumber
		\\
	&
	=
		2
		\delta_{\vec p_\final, \vec p_\init + \hbar \vec k - \hbar \vec k' }
		\delta_{\vec k_\final, \vec k'}
		\delta_{\vec k, \vec k_\init}
		,
	\end{align}
	and $(***)$ becomes
	\begin{align}
		(***)
	&
	=
		\inprod{\vec p_\final} {\vec p_\init}~
		\vacsonicbra
		\op a_{\vec k_\final}
		\op a^\dagger_{\vec k_\init}
		\vacsonicket
		\delta_{\vec k, \vec k'}
		\nonumber
		\\
	&
	=
		\delta_{\vec p_\final, \vec p_\init}
		\delta_{\vec k_\final, \vec k_\init}
		\delta_{\vec k, \vec k'}
		.
	\end{align}
	The amplitude Eq.~\eqref{Eq:PAD} is therefore
	\begin{align}
		\bra \final {\op H_\text{int} } \ket \init
	=
		\frac{\charge}{2}
		\frac{\hbar c_\sonic}{L^3}
		\sum_{\vec k, \vec k'}
		\frac{1}{2\sqrt{k k'}}
		\left[
			\underbrace{
				2
				\delta_{\vec p_\final, \vec p_\init + \hbar \vec k - \hbar \vec k' }
				\delta_{\vec k_\final, \vec k'}
				\delta_{\vec k, \vec k_\init}
			}_{(1)}
		+
			\underbrace{
				\delta_{\vec p_\final, \vec p_\init}
				\delta_{\vec k_\final, \vec k_\init}
				\delta_{\vec k, \vec k'}
			}_{(2)}
		\right]
		.
	\end{align}
	Evaluating $(2)$ first we obtain
	\begin{align}
		(2)
	&
	=
		\frac{\charge}{2}
		\frac{\hbar c_\sonic}{L^3}
		\delta_{\vec p_\final, \vec p_\init}
		\delta_{\vec k_\final, \vec k_\init}
		\sum_{\vec k, \vec k'}
			\frac{1}{2\sqrt{k k'}}
			\delta_{\vec k, \vec k'}
		\nonumber
		\\
	&
	=
	 	\frac{\charge}{2}
		\frac{\hbar c_\sonic}{L^3}
		\delta_{\vec p_\final, \vec p_\init}
		\delta_{\vec k_\final, \vec k_\init}
		\sum_{\vec k}
			\frac{1}{2 k}
		.
	\end{align}
	This divergent term represents the case of no scattering.
	Since we are only interested in the set $\mathcal{S}$ of final states that \textit{do not} contain the initial states (i.e.,~we are excluding the case of no scattering), we will not consider this term. Term~$(1)$ instead is
	\begin{align}
		(1)
	&
	=
		\frac{\charge}{2}
		\frac{\hbar c_\sonic}{L^3}
		\sum_{\vec k, \vec k'}
		\frac{1}{2\sqrt{k k'}}
			2
			\delta_{\vec p_\final, \vec p_\init + \hbar \vec k - \hbar \vec k' }
			\delta_{\vec k_\final, \vec k'}
			\delta_{\vec k, \vec k_\init}
		\nonumber
		\\
	&
	=
		\frac{\charge}{2}
		\frac{\hbar c_\sonic}{L^3}
		\frac{1}{\sqrt{k_\init k_\final}}
				\delta_{\vec p_\final, \vec p_\init + \hbar \vec k_\init - \hbar \vec k_\final }
		.
	\end{align}
	The amplitude is therefore
	\begin{align}
	\label{Eq:ProbAmpDisc}
		\bra \final {\op H_\text{int} } \ket \init
	=
		\frac{\charge}{2}
		\frac{\hbar c_\sonic}{L^3}
		\frac{1}{\sqrt{k_\init k_\final}}
				\delta_{\vec p_\final, \vec p_\init + \hbar \vec k_\init - \hbar \vec k_\final }
		.
	\end{align}
	The transition rate~\cite{Sakurai:1167961} for an initial state $\ket \init$ at time $-\infty$ and a final state $\ket \final$ at time $T$ is
	\begin{align}
		w_{\init \rightarrow \final}
	=
		\frac {2 \pi} {\hbar}
		\abss{ \bra \final {\op H_\text{int} }\ket \init }^2
		\delta(E_{\final \init})
		,
		\label{Eq:AppendixTransitionRatePenultimate}
	\end{align}
	where the states and the interaction Hamiltonian are defined in Schr\"odinger picture. Substituting Eq.~\eqref{Eq:ProbAmpDisc} into Eq.~\eqref{Eq:AppendixTransitionRatePenultimate} gives the explicit form of the transition rate:
	\begin{align}
		w_{\init \rightarrow \final}
	&
	=
		\frac{2 \pi}{\hbar}
		\abs{
			\frac{\charge}{2}
			\frac{\hbar c_\sonic}{L^3}
			\frac{1}{\sqrt{k_\init k_\final}}
			\delta_{\vec p_\final, \vec p_\init + \hbar \vec k_\init \hbar \vec k_\final }
			}^2
			\delta(E_{\final \init})
			\nonumber
			\\*
	&
	=
		\frac{2 \pi}{\hbar}
		\frac{\charge^2}{4}
			\frac{(\hbar c_\sonic)^2}{L^6}
		\frac{1}{k_\init k_\final}
		\left(
			\delta_{\vec p_\final, \vec p_\init + \hbar \vec k_\init - \hbar  \vec k_\final}
		\right)^2
		\delta(E_{\final \init})
		\nonumber
		\\*
	&
	=
		\frac{2 \pi}{\hbar}
		\frac{\charge^2}{4}
			\frac{(\hbar c_\sonic)^2}{L^6}
		\frac{1}{k_\init k_\final}
		\delta_{\vec p_\final, \vec p_\init + \hbar \vec k_\init - \hbar  \vec k_\final}
		\delta(E_{\final \init})
		.
	\end{align}

	\end{widetext}

\bibliography{Particle_scattering_in_analogue-gravity_models_CHANGED}

\end{document}